\documentclass[aps,prd,preprint,superscriptaddress]{revtex4-2}
\usepackage[T1]{fontenc}
\usepackage{lmodern}
\usepackage[normalem]{ulem}
\usepackage{amsmath,amssymb,bm,mathrsfs,amsthm,mathtools}
\usepackage{graphicx}
\usepackage{xcolor}
\usepackage{tikz}
\usetikzlibrary{arrows.meta,positioning,fit,calc,shadows,decorations.pathmorphing}

\usepackage[colorlinks=true,allcolors=blue]{hyperref}
\newcommand{\dd}{\mathrm{d}}

\definecolor{Gold}{RGB}{181,137,0}

\usepackage{booktabs,microtype,xcolor,needspace}
\usepackage{pgfplots}
\usepgfplotslibrary{groupplots}
\pgfplotsset{compat=1.18}
\newif\ifshowchanges\showchangestrue
\definecolor{revisionred}{rgb}{0.85,0,0}

\newenvironment{revblock}{\begingroup}{\endgroup}

\newtagform{revised}{\color{black}(}{)}
\usetagform{revised}
\newtheorem{theorem}{Theorem}
\newtheorem{proposition}{Proposition}
\newtheorem{lemma}{Lemma}
\newcommand{\Id}{\mathrm{Id}}
\newcommand{\Ge}{\widetilde G}
\makeatletter
\expandafter\def\csname changedref@s2:collinear\endcsname{1}
\expandafter\def\csname changedref@s2:translations\endcsname{1}
\expandafter\def\csname changedref@s2:convolution\endcsname{1}
\expandafter\def\csname changedref@s2:addsupport\endcsname{1}
\expandafter\def\csname changedref@s2:star\endcsname{1}
\expandafter\def\csname changedref@s2:leibniz\endcsname{1}
\expandafter\def\csname changedref@s2:starsupport\endcsname{1}
\expandafter\def\csname changedref@s2:quarticderivation\endcsname{1}
\expandafter\def\csname changedref@s2:phase\endcsname{1}
\expandafter\def\csname changedref@s2:collinearphase\endcsname{1}
\expandafter\def\csname changedref@s3:shells\endcsname{1}
\expandafter\def\csname changedref@s3:conservation\endcsname{1}
\expandafter\def\csname changedref@s3:angular\endcsname{1}
\expandafter\def\csname changedref@s3:y0\endcsname{1}
\expandafter\def\csname changedref@s3:fixedenergy\endcsname{1}
\expandafter\def\csname changedref@s3:deltap\endcsname{1}
\expandafter\def\csname changedref@s3:source\endcsname{1}
\expandafter\def\csname changedref@s3:Cgamma\endcsname{1}
\expandafter\def\csname changedref@s3:compensation\endcsname{1}
\expandafter\def\csname changedref@s3:separation\endcsname{1}
\expandafter\def\csname changedref@s3:response\endcsname{1}
\expandafter\def\csname changedref@s3:equalorder\endcsname{1}
\expandafter\def\csname changedref@s3:geometry\endcsname{1}
\expandafter\def\csname changedref@s3:crossedge\endcsname{1}
\expandafter\def\csname changedref@s3:crosssource\endcsname{1}
\expandafter\def\csname changedref@s3:scaling\endcsname{1}
\expandafter\def\csname changedref@s3:threshold\endcsname{1}
\expandafter\def\csname changedref@s3:bound\endcsname{1}
\expandafter\def\csname changedref@s3:boundnumbers\endcsname{1}
\expandafter\def\csname changedref@s3:benchmark\endcsname{1}
\expandafter\def\csname changedref@s3:finalnumbers\endcsname{1}
\makeatother
\allowdisplaybreaks
\newcommand{\comptonplot}{%
\begin{tikzpicture}
\begin{groupplot}[group style={group size=2 by 1,horizontal sep=1.8cm},width=.42\linewidth,height=5.0cm,xmode=log,ymode=log,xmin=1,xmax=1000,xlabel={$E_e\ (\mathrm{GeV})$},tick label style={font=\small},label style={font=\small},title style={font=\small},minor tick num=9,grid=major,grid style={gray!20},legend style={draw=none,font=\scriptsize,cells={anchor=west}}]
\nextgroupplot[title={(a)},ylabel={$|C_{\rm op}|=|C_\gamma|\ (\mathrm{GeV}^3)$},ymin=1e5,ymax=2e18,ytick={1e6,1e9,1e12,1e15,1e18},legend pos=north west]
\addplot[thick,blue,solid] coordinates {(1,1.9106895216e+05) (1.071519305,2.8774823018e+05) (1.148153621,4.3319405636e+05) (1.230268771,6.5191222329e+05) (1.318256739,9.8066662820e+05) (1.412537545,1.4745768477e+06) (1.513561248,2.2162283258e+06) (1.621810097,3.3292713278e+06) (1.737800829,4.9987008047e+06) (1.862087137,7.5010682799e+06) (1.995262315,1.1249440565e+07) (2.13796209,1.6860225132e+07) (2.290867653,2.5252373519e+07) (2.454708916,3.7794454543e+07) (2.630267992,5.6522387432e+07) (2.818382931,8.4461295958e+07) (3.01995172,1.2610050976e+08) (3.235936569,1.8809338489e+08) (3.467368505,2.8028647399e+08) (3.715352291,4.1723011498e+08) (3.981071706,6.2039107211e+08) (4.265795188,9.2138643010e+08) (4.570881896,1.3666991436e+09) (4.897788194,2.0245371685e+09) (5.248074602,2.9947845822e+09) (5.623413252,4.4233986350e+09) (6.025595861,6.5231781789e+09) (6.45654229,9.6036304854e+09) (6.918309709,1.4113782040e+10) (7.413102413,2.0703331517e+10) (7.943282347,3.0309685979e+10) (8.511380382,4.4281361060e+10) (9.120108394,6.4552233099e+10) (9.77237221,9.3886556929e+10) (10.47128548,1.3622195704e+11) (11.22018454,1.9714733199e+11) (12.02264435,2.8456549454e+11) (12.88249552,4.0960728008e+11) (13.80384265,5.8788585875e+11) (14.79108388,8.4120835069e+11) (15.84893192,1.1998980556e+12) (16.98243652,1.7059263680e+12) (18.19700859,2.4171106648e+12) (19.498446,3.4127051975e+12) (20.89296131,4.8007984924e+12) (22.38721139,6.7280351938e+12) (23.98832919,9.3923048599e+12) (25.70395783,1.3059186950e+13) (27.54228703,1.8083111852e+13) (29.51209227,2.4934393593e+13) (31.6227766,3.4233511744e+13) (33.88441561,4.6794268168e+13) (36.30780548,6.3677718683e+13) (38.9045145,8.6259079700e+13) (41.68693835,1.1631013505e+14) (44.66835922,1.5610001807e+14) (47.86300923,2.0851761929e+14) (51.2861384,2.7721927221e+14) (54.95408739,3.6680580324e+14) (58.88436554,4.8303350261e+14) (63.09573445,6.3306409174e+14) (67.60829754,8.2575934165e+14) (72.44359601,1.0720266554e+15) (77.62471166,1.3852226860e+15) (83.17637711,1.7816229460e+15) (89.12509381,2.2809664066e+15) (95.4992586,2.9070853148e+15) (102.3292992,3.6886319150e+15) (109.6478196,4.6599154834e+15) (117.4897555,5.8618651171e+15) (125.8925412,7.3431361032e+15) (134.8962883,9.1613804846e+15) (144.5439771,1.1384705680e+16) (154.8816619,1.4093348777e+16) (165.9586907,1.7381598457e+16) (177.827941,2.1360001512e+16) (190.5460718,2.6157896630e+16) (204.1737945,3.1926324745e+16) (218.7761624,3.8841372730e+16) (234.4228815,4.7108015899e+16) (251.1886432,5.6964534604e+16) (269.1534804,6.8687591579e+16) (288.4031503,8.2598069589e+16) (309.0295433,9.9067783812e+16) (331.1311215,1.1852720041e+17) (354.8133892,1.4147431220e+17) (380.1893963,1.6848484481e+17) (407.3802778,2.0022399233e+17) (436.5158322,2.3745991095e+17) (467.7351413,2.8107923298e+17) (501.1872336,3.3210490256e+17) (537.0317964,3.9171667879e+17) (575.4399373,4.6127470364e+17) (616.5950019,5.4234659049e+17) (660.693448,6.3673855705e+17) (707.9457844,7.4653120388e+17) (758.577575,8.7412062903e+17) (812.8305162,1.0222656715e+18) (870.96359,1.1941421943e+18) (933.2543008,1.3934054519e+18) (1000,1.6242617436e+18)};
\addlegendentry{$\lambda=1064\ \mathrm{nm}$}
\addplot[thick,red,dashed] coordinates {(1,3.5646664521e+05) (1.071519305,5.3424638577e+05) (1.148153621,8.0014745711e+05) (1.230268771,1.1975256411e+06) (1.318256739,1.7908756194e+06) (1.412537545,2.6760245001e+06) (1.513561248,3.9951748517e+06) (1.621810097,5.9590662928e+06) (1.737800829,8.8795646692e+06) (1.862087137,1.3217492299e+07) (1.995262315,1.9652682415e+07) (2.13796209,2.9186359573e+07) (2.290867653,4.3290414864e+07) (2.454708916,6.4124519328e+07) (2.630267992,9.4851079589e+07) (2.818382931,1.4009086388e+08) (3.01995172,2.0658019611e+08) (3.235936569,3.0411595514e+08) (3.467368505,4.4690997063e+08) (3.715352291,6.5552347066e+08) (3.981071706,9.5961993792e+08) (4.265795188,1.4018675892e+09) (4.570881896,2.0434492435e+09) (4.897788194,2.9718086480e+09) (5.248074602,4.3114925699e+09) (5.623413252,6.2392550728e+09) (6.025595861,9.0049968121e+09) (6.45654229,1.2960645505e+10) (6.918309709,1.8599777448e+10) (7.413102413,2.6611673936e+10) (7.943282347,3.7954647390e+10) (8.511380382,5.3954913265e+10) (9.120108394,7.6439085206e+10) (9.77237221,1.0791059746e+11) (10.47128548,1.5178307881e+11) (11.22018454,2.1268698646e+11) (12.02264435,2.9686972375e+11) (12.88249552,4.1271407598e+11) (13.80384265,5.7140515626e+11) (14.79108388,7.8778220003e+11) (15.84893192,1.0814185074e+12) (16.98243652,1.4779806165e+12) (18.19700859,2.0109263934e+12) (19.498446,2.7236111266e+12) (20.89296131,3.6718809005e+12) (22.38721139,4.9272434861e+12) (23.98832919,6.5807187344e+12) (25.70395783,8.7474830655e+12) (27.54228703,1.1572436225e+13) (29.51209227,1.5236833237e+13) (31.6227766,1.9966140754e+13) (33.88441561,2.6039295162e+13) (36.30780548,3.3799560494e+13) (38.9045145,4.3667207997e+13) (41.68693835,5.6154267026e+13) (44.66835922,7.1881629661e+13) (47.86300923,9.1598830130e+13) (51.2861384,1.1620686592e+14) (54.95408739,1.4678448166e+14) (58.88436554,1.8461840059e+14) (63.09573445,2.3123806361e+14) (67.60829754,2.8845552317e+14) (72.44359601,3.5841124158e+14) (77.62471166,4.4362666090e+14) (83.17637711,5.4706454846e+14) (89.12509381,6.7219827823e+14) (95.4992586,8.2309138906e+14) (102.3292992,1.0044889670e+15) (109.6478196,1.2219226355e+15) (117.4897555,1.4818312086e+15) (125.8925412,1.7916993714e+15) (134.8962883,2.1602171087e+15) (144.5439771,2.5974630082e+15) (154.8816619,3.1151150304e+15) (165.9586907,3.7266928734e+15) (177.827941,4.4478366714e+15) (190.5460718,5.2966274699e+15) (204.1737945,6.2939557266e+15) (218.7761624,7.4639450131e+15) (234.4228815,8.8344391564e+15) (251.1886432,1.0437562277e+16) (269.1534804,1.2310362586e+16) (288.4031503,1.4495552412e+16) (309.0295433,1.7042358774e+16) (331.1311215,2.0007500938e+16) (354.8133892,2.3456313854e+16) (380.1893963,2.7464039128e+16) (407.3802778,3.2117308442e+16) (436.5158322,3.7515848002e+16) (467.7351413,4.3774436839e+16) (501.1872336,5.1025156662e+16) (537.0317964,5.9419976540e+16) (575.4399373,6.9133722132e+16) (616.5950019,8.0367486524e+16) (660.693448,9.3352548226e+16) (707.9457844,1.0835487157e+17) (758.577575,1.2568027597e+17) (812.8305162,1.4568037318e+17) (870.96359,1.6875938661e+17) (933.2543008,1.9538198346e+17) (1000,2.2608226988e+17)};
\addlegendentry{$\lambda=532\ \mathrm{nm}$}
\addplot[thick,green!60!black,densely dotted] coordinates {(1,6.2255453314e+05) (1.071519305,9.2452208826e+05) (1.148153621,1.3712283845e+06) (1.230268771,2.0310549981e+06) (1.318256739,3.0041258936e+06) (1.412537545,4.4367266493e+06) (1.513561248,6.5421005662e+06) (1.621810097,9.6303487424e+06) (1.737800829,1.4151278628e+07) (1.862087137,2.0755596005e+07) (1.995262315,3.0381974303e+07) (2.13796209,4.4380468309e+07) (2.290867653,6.4686735802e+07) (2.454708916,9.4066939190e+07) (2.630267992,1.3646046658e+08) (2.818382931,1.9745730307e+08) (3.01995172,2.8495970431e+08) (3.235936569,4.1009464443e+08) (3.467368505,5.8846538189e+08) (3.715352291,8.4185866376e+08) (3.981071706,1.2005600363e+09) (4.265795188,1.7064751246e+09) (4.570881896,2.4173114583e+09) (4.897788194,3.4121454942e+09) (5.248074602,4.7987850639e+09) (5.623413252,6.7234407440e+09) (6.025595861,9.3833427268e+09) (6.45654229,1.3043084609e+10) (6.918309709,1.8055643779e+10) (7.413102413,2.4889221033e+10) (7.943282347,3.4161260429e+10) (8.511380382,4.6681254563e+10) (9.120108394,6.3504210080e+10) (9.77237221,8.5996942977e+10) (10.47128548,1.1591969239e+11) (11.22018454,1.5552588495e+11) (12.02264435,2.0768324988e+11) (12.88249552,2.7601987988e+11) (13.80384265,3.6509925831e+11) (14.79108388,4.8062873589e+11) (15.84893192,6.2970645017e+11) (16.98243652,8.2111225138e+11) (18.19700859,1.0656488471e+12) (19.498446,1.3765401262e+12) (20.89296131,1.7698944973e+12) (22.38721139,2.2652421039e+12) (23.98832919,2.8861559958e+12) (25.70395783,3.6609687754e+12) (27.54228703,4.6235979408e+12) (29.51209227,5.8144951517e+12) (31.6227766,7.2817370018e+12) (33.88441561,9.0822776314e+12) (36.30780548,1.1283386716e+13) (38.9045145,1.3964300075e+13) (41.68693835,1.7218114422e+13) (44.66835922,2.1153962704e+13) (47.86300923,2.5899512149e+13) (51.2861384,3.1603833585e+13) (54.95408739,3.8440698077e+13) (58.88436554,4.6612365394e+13) (63.09573445,5.6353938564e+13) (67.60829754,6.7938369932e+13) (72.44359601,8.1682216899e+13) (77.62471166,9.7952260151e+13) (83.17637711,1.1717311396e+14) (89.12509381,1.3983597739e+14) (95.4992586,1.6650869727e+14) (102.3292992,1.9784733920e+14) (109.6478196,2.3460949178e+14) (117.4897555,2.7766956292e+14) (125.8925412,3.2803636495e+14) (134.8962883,3.8687332981e+14) (144.5439771,4.5552174566e+14) (154.8816619,5.3552746462e+14) (165.9586907,6.2867159778e+14) (177.827941,7.3700579045e+14) (190.5460718,8.6289275804e+14) (204.1737945,1.0090528647e+15) (218.7761624,1.1786176415e+15) (234.4228815,1.3751912763e+15) (251.1886432,1.6029212568e+15) (269.1534804,1.8665795277e+15) (288.4031503,2.1716557221e+15) (309.0295433,2.5244642585e+15) (331.1311215,2.9322673624e+15) (354.8133892,3.4034163744e+15) (380.1893963,3.9475140593e+15) (407.3802778,4.5756010309e+15) (436.5158322,5.3003698705e+15) (467.7351413,6.1364110475e+15) (501.1872336,7.1004953583e+15) (537.0317964,8.2118983001e+15) (575.4399373,9.4927725996e+15) (616.5950019,1.0968576036e+16) (660.693448,1.2668562759e+16) (707.9457844,1.4626347516e+16) (758.577575,1.6880553600e+16) (812.8305162,1.9475556925e+16) (870.96359,2.2462340496e+16) (933.2543008,2.5899475610e+16) (1000,2.9854248616e+16)};
\addlegendentry{$\lambda=266\ \mathrm{nm}$}
\addplot[only marks,black,mark=*] coordinates {(11,193140981944.18085)};
\node[font=\scriptsize,anchor=north west] at (axis cs:14,1e11) {Letter benchmark};
\nextgroupplot[title={(b)},ylabel={$|\Delta y|_{\max}/y_0$},ymin=1e-24,ymax=3e-15,ytick={1e-23,1e-21,1e-19,1e-17,1e-15}]
\addplot[thick,blue,solid] coordinates {(1,8.9884195263e-24) (1.071519305,1.1804571825e-23) (1.148153621,1.5498961879e-23) (1.230268771,2.0343823825e-23) (1.318256739,2.6695114666e-23) (1.412537545,3.5017997009e-23) (1.513561248,4.5919972492e-23) (1.621810097,6.0193921308e-23) (1.737800829,7.8873951764e-23) (1.862087137,1.0330779268e-22) (1.995262315,1.3525051810e-22) (2.13796209,1.7698573769e-22) (2.290867653,2.3148209099e-22) (2.454708916,3.0259503983e-22) (2.630267992,3.9532667210e-22) (2.818382931,5.1615964736e-22) (3.01995172,6.7348569404e-22) (3.235936569,8.7815440637e-22) (3.467368505,1.1441747213e-21) (3.715352291,1.4896096566e-21) (3.981071706,1.9377149854e-21) (4.265795188,2.5183848742e-21) (4.570881896,3.2699825501e-21) (4.897788194,4.2416522402e-21) (5.248074602,5.4962304629e-21) (5.623413252,7.1139007712e-21) (6.025595861,9.1967668016e-21) (6.45654229,1.1874554486e-20) (6.918309709,1.5311696006e-20) (7.413102413,1.9716095847e-20) (7.943282347,2.5349933309e-20) (8.511380382,3.2542915933e-20) (9.120108394,4.1708464011e-20) (9.77237221,5.3363376708e-20) (10.47128548,6.8151603713e-20) (11.22018454,8.6872820408e-20) (12.02264435,1.1051657615e-19) (12.88249552,1.4030285019e-19) (13.80384265,1.7772990292e-19) (14.79108388,2.2463034541e-19) (15.84893192,2.8323636079e-19) (16.98243652,3.5625498904e-19) (18.19700859,4.4695432526e-19) (19.498446,5.5926137211e-19) (20.89296131,6.9787212465e-19) (22.38721139,8.6837424518e-19) (23.98832919,1.0773824057e-18) (25.70395783,1.3326860379e-18) (27.54228703,1.6434088444e-18) (29.51209227,2.0201789961e-18) (31.6227766,2.4753084971e-18) (33.88441561,3.0229797512e-18) (36.30780548,3.6794369509e-18) (38.9045145,4.4631795523e-18) (41.68693835,5.3951548410e-18) (44.66835922,6.4989464498e-18) (47.86300923,7.8009557033e-18) (51.2861384,9.3305728434e-18) (54.95408739,1.1120335549e-17) (58.88436554,1.3206072705e-17) (63.09573445,1.5627032084e-17) (67.60829754,1.8425991441e-17) (72.44359601,2.1649353482e-17) (77.62471166,2.5347226174e-17) (83.17637711,2.9573490848e-17) (89.12509381,3.4385861560e-17) (95.4992586,3.9845939979e-17) (102.3292992,4.6019270841e-17) (109.6478196,5.2975403520e-17) (117.4897555,6.0787965627e-17) (125.8925412,6.9534754688e-17) (134.8962883,7.9297853886e-17) (144.5439771,9.0163777619e-17) (154.8816619,1.0222365224e-16) (165.9586907,1.1557343681e-16) (177.827941,1.3031418817e-16) (190.5460718,1.4655237398e-16) (204.1737945,1.6440023674e-16) (218.7761624,1.8397621139e-16) (234.4228815,2.0540539830e-16) (251.1886432,2.2882009346e-16) (269.1534804,2.5436037705e-16) (288.4031503,2.8217476147e-16) (309.0295433,3.1242090011e-16) (331.1311215,3.4526635776e-16) (354.8133892,3.8088944411e-16) (380.1893963,4.1948011187e-16) (407.3802778,4.6124092133e-16) (436.5158322,5.0638807390e-16) (467.7351413,5.5515251712e-16) (501.1872336,6.0778112479e-16) (537.0317964,6.6453795584e-16) (575.4399373,7.2570559661e-16) (616.5950019,7.9158659149e-16) (660.693448,8.6250496778e-16) (707.9457844,9.3880786099e-16) (758.577575,1.0208672478e-15) (812.8305162,1.1090817941e-15) (870.96359,1.2038788274e-15) (933.2543008,1.3057164409e-15) (1000,1.4150857417e-15)};
\addplot[thick,red,dashed] coordinates {(1,8.5316380043e-24) (1.071519305,1.1164118577e-23) (1.148153621,1.4601442686e-23) (1.230268771,1.9086744589e-23) (1.318256739,2.4935455975e-23) (1.412537545,3.2556357600e-23) (1.513561248,4.2478589706e-23) (1.621810097,5.5386243271e-23) (1.737800829,7.2162571727e-23) (1.862087137,9.3946379048e-23) (1.995262315,1.2220377541e-22) (2.13796209,1.5881926903e-22) (2.290867653,2.0621110870e-22) (2.454708916,2.6747693527e-22) (2.630267992,3.4657717325e-22) (2.818382931,4.4856522976e-22) (3.01995172,5.7987550067e-22) (3.235936569,7.4868244620e-22) (3.467368505,9.6534661811e-22) (3.715352291,1.2429665205e-21) (3.981071706,1.5980585739e-21) (4.265795188,2.0513912222e-21) (4.570881896,2.6290033392e-21) (4.897788194,3.3634414981e-21) (5.248074602,4.2952552574e-21) (5.623413252,5.4747942465e-21) (6.025595861,6.9643552964e-21) (6.45654229,8.8407319030e-21) (6.918309709,1.1198221601e-20) (7.413102413,1.4152148981e-20) (7.943282347,1.7842962692e-20) (8.511380382,2.2440963310e-20) (9.120108394,2.8151715011e-20) (9.77237221,3.5222187077e-20) (10.47128548,4.3947660927e-20) (11.22018454,5.4679424507e-20) (12.02264435,6.7833258218e-20) (12.88249552,8.3898695358e-20) (13.80384265,1.0344901574e-19) (14.79108388,1.2715190448e-19) (15.84893192,1.5578067992e-19) (16.98243652,1.9022596699e-19) (18.19700859,2.3150766645e-19) (19.498446,2.8078704829e-19) (20.89296131,3.3937878171e-19) (22.38721139,4.0876270525e-19) (23.98832919,4.9059514212e-19) (25.70395783,5.8671957698e-19) (27.54228703,6.9917653372e-19) (29.51209227,8.3021252765e-19) (31.6227766,9.8228801002e-19) (33.88441561,1.1580842762e-18) (36.30780548,1.3605093679e-18) (38.9045145,1.5927030631e-18) (41.68693835,1.8580411104e-18) (44.66835922,2.1601389243e-18) (47.86300923,2.5028550115e-18) (51.2861384,2.8902944443e-18) (54.95408739,3.3268127280e-18) (58.88436554,3.8170204338e-18) (63.09573445,4.3657889736e-18) (67.60829754,4.9782578926e-18) (72.44359601,5.6598440364e-18) (77.62471166,6.4162529280e-18) (83.17637711,7.2534926569e-18) (89.12509381,8.1778905444e-18) (95.4992586,9.1961128147e-18) (102.3292992,1.0315187461e-17) (109.6478196,1.1542530457e-17) (117.4897555,1.2885975449e-17) (125.8925412,1.4353807008e-17) (134.8962883,1.5954797546e-17) (144.5439771,1.7698247948e-17) (154.8816619,1.9594031998e-17) (165.9586907,2.1652644665e-17) (177.827941,2.3885254335e-17) (190.5460718,2.6303759083e-17) (204.1737945,2.8920847102e-17) (218.7761624,3.1750061452e-17) (234.4228815,3.4805869279e-17) (251.1886432,3.8103735738e-17) (269.1534804,4.1660202849e-17) (288.4031503,4.5492973568e-17) (309.0295433,4.9621001405e-17) (331.1311215,5.4064585930e-17) (354.8133892,5.8845474582e-17) (380.1893963,6.3986971216e-17) (407.3802778,6.9514051870e-17) (436.5158322,7.5453488289e-17) (467.7351413,8.1833979769e-17) (501.1872336,8.8686293945e-17) (537.0317964,9.6043417195e-17) (575.4399373,1.0394071537e-16) (616.5950019,1.1241610565e-16) (660.693448,1.2151024029e-16) (707.9457844,1.3126670322e-16) (758.577575,1.4173222042e-16) (812.8305162,1.5295688506e-16) (870.96359,1.6499439851e-16) (933.2543008,1.7790232844e-16) (1000,1.9174238514e-16)};
\addplot[thick,green!60!black,densely dotted] coordinates {(1,7.7068835743e-24) (1.071519305,1.0015755570e-23) (1.148153621,1.3004033948e-23) (1.230268771,1.6866967630e-23) (1.318256739,2.1854171420e-23) (1.412537545,2.8284125336e-23) (1.513561248,3.6562312886e-23) (1.621810097,4.7203832416e-23) (1.737800829,6.0861480204e-23) (1.862087137,7.8360492295e-23) (1.995262315,1.0074134463e-22) (2.13796209,1.2931224742e-22) (2.290867653,1.6571322779e-22) (2.454708916,2.1199397052e-22) (2.630267992,2.7070787408e-22) (2.818382931,3.4502506801e-22) (3.01995172,4.3886741649e-22) (3.235936569,5.5706878378e-22) (3.467368505,7.0556404118e-22) (3.715352291,8.9161042715e-22) (3.981071706,1.1240449054e-21) (4.265795188,1.4135810712e-21) (4.570881896,1.7731489026e-21) (4.897788194,2.2182802143e-21) (5.248074602,2.7675420196e-21) (5.623413252,3.4430191315e-21) (6.025595861,4.2708462202e-21) (6.45654229,5.2817882164e-21) (6.918309709,6.5118664204e-21) (7.413102413,8.0030260061e-21) (7.943282347,9.8038388609e-21) (8.511380382,1.1970233970e-20) (9.120108394,1.4566245942e-20) (9.77237221,1.7664770916e-20) (10.47128548,2.1348318072e-20) (11.22018454,2.5709744500e-20) (12.02264435,3.0852961211e-20) (12.88249552,3.6893598887e-20) (13.80384265,4.3959623367e-20) (14.79108388,5.2191893062e-20) (15.84893192,6.1744653282e-20) (16.98243652,7.2785965775e-20) (18.19700859,8.5498075547e-20) (19.498446,1.0007772092e-19) (20.89296131,1.1673639679e-19) (22.38721139,1.3570058466e-19) (23.98832919,1.5721196664e-19) (25.70395783,1.8152764290e-19) (27.54228703,2.0892037472e-19) (29.51209227,2.3967887605e-19) (31.6227766,2.7410817734e-19) (33.88441561,3.1253008497e-19) (36.30780548,3.5528375860e-19) (38.9045145,4.0272642735e-19) (41.68693835,4.5523426351e-19) (44.66835922,5.1320343041e-19) (47.86300923,5.7705131848e-19) (51.2861384,6.4721798144e-19) (54.95408739,7.2416778196e-19) (58.88436554,8.0839125468e-19) (63.09573445,9.0040719261e-19) (67.60829754,1.0007649620e-18) (72.44359601,1.1100470502e-18) (77.62471166,1.2288718502e-18) (83.17637711,1.3578966871e-18) (89.12509381,1.4978210910e-18) (95.4992586,1.6493903229e-18) (102.3292992,1.8133991606e-18) (109.6478196,1.9906959550e-18) (117.4897555,2.1821869659e-18) (125.8925412,2.3888409928e-18) (134.8962883,2.6116943139e-18) (144.5439771,2.8518559528e-18) (154.8816619,3.1105132910e-18) (165.9586907,3.3889380507e-18) (177.827941,3.6884926717e-18) (190.5460718,4.0106371105e-18) (204.1737945,4.3569360920e-18) (218.7761624,4.7290668467e-18) (234.4228815,5.1288273696e-18) (251.1886432,5.5581452397e-18) (269.1534804,6.0190870411e-18) (288.4031503,6.5138684323e-18) (309.0295433,7.0448649107e-18) (331.1311215,7.6146233248e-18) (354.8133892,8.2258741899e-18) (380.1893963,8.8815448661e-18) (407.3802778,9.5847736633e-18) (436.5158322,1.0338924940e-17) (467.7351413,1.1147605272e-17) (501.1872336,1.2014680763e-17) (537.0317964,1.2944295587e-17) (575.4399373,1.3940891849e-17) (616.5950019,1.5009230859e-17) (660.693448,1.6154415924e-17) (707.9457844,1.7381916762e-17) (758.577575,1.8697595663e-17) (812.8305162,2.0107735517e-17) (870.96359,2.1619069845e-17) (933.2543008,2.3238814976e-17) (1000,2.4974704525e-17)};
\addplot[only marks,black,mark=*] coordinates {(11,5.137214862862807e-20)};
\end{groupplot}\end{tikzpicture}}

\begin{document}

\title{One-Particle Equivalence of Lorentz-Violating and Lorentz-Deformed Theories}

\author{Arshid Shabir}
\email{aslone186@gmail.com}
\affiliation{Canadian Quantum Research Center, 460 Doyle Ave 106, Kelowna, BC V1Y 0C2, Canada}

\author{Francesco Marino}
\email{francesco.marino@ino.cnr.it}
\affiliation{CNR-Istituto Nazionale di Ottica, Via N. Carrara 1, I-50019 Sesto Fiorentino (FI), Italy}
\affiliation{INFN, Via Sansone 1, I-50019 Sesto Fiorentino (FI), Italy}

\author{Mir Faizal}
\email{mirfaizalmir@gmail.com}
\affiliation{Canadian Quantum Research Center, 460 Doyle Ave 106, Kelowna, BC V1Y 0C2, Canada}
\affiliation{Irving K. Barber School of Arts and Sciences, University of British Columbia Okanagan, Kelowna, BC V1V 1V7, Canada}
\affiliation{Department of Mathematical Sciences, Durham University, Upper Mountjoy, Stockton Road, Durham DH1 3LE, United Kingdom}
\affiliation{Computational Mathematics Group, Faculty of Sciences, Hasselt University, Agoralaan Gebouw D, Diepenbeek, 3590 Belgium}

%\date{\today}

\begin{abstract}

Planck-scale departures from special relativity are commonly described by theories that break Lorentz symmetry and theories in which such symmetry is deformed. We prove that, under a global shell-linearizability condition, these two descriptions are physically equivalent for a single particle, corresponding to unitarily equivalent realizations of the same Wigner orbit expressed in different momentum coordinates. This equivalence breaks down for multiparticle systems, except when the momentum-composition law can be reduced to the ordinary addition of momenta. Our results imply that the distinction between Lorentz violating and Lorentz-deformed kinematics can only be tested by genuine multiparticle observables. We discuss Compton edge measurements as a prototype example of such a multiparticle probe.

\end{abstract}

\maketitle

\newpage

\section{Introduction}

A widely shared expectation in quantum gravity is that the classical description of spacetime as a smooth continuous manifold should eventually break down near the Planck scale. Several approaches, including string-inspired models \cite{KosteleckySamuel1991,KosteleckySamuel1989}, loop-quantum-gravity-inspired scenarios \cite{GambiniPullin1999,Alfaro2000}, causal-set and discrete-spacetime models \cite{Bombelli1987,Snyder1947}, noncommutative and $\kappa$-Minkowski geometries \cite{MajidRuegg1994,Lukierski1995}, relative locality \cite{RelativeLocality2011,RelativeLocalityDeepening2011}, and holographic descriptions of emergent spacetime \cite{Qi2018,Dong2016}, suggest that ordinary Lorentz invariance may emerge as an effective symmetry at low energies. This possibility raises fundamental questions about the fate of Lorentz symmetry in the Planck regime.

Early investigations focused on explicit Lorentz invariance violation (LIV) \cite{Biller1999,Kifune1999,Aloisio2000,ProtheroeMeyer2000,AmelinoPiran2001}. In these scenarios, Lorentz transformations retain their standard special-relativistic form, while the fundamental dynamical laws are no longer invariant under them, leading to the emergence of a preferred class of inertial observers \cite{Mattingly2005,Liberati2013}. A different perspective was introduced with the proposal of doubly special relativity (DSR) \cite{Amelino2001,Amelino2002,MagueijoSmolin2002,KowalskiNowak2002,JudesVisser2003}. In DSR the relativity principle is preserved, but the action of Lorentz transformations is deformed so as to incorporate, in addition to the invariant speed scale $c$, a second observer-independent scale associated with the Planck length $L_p$ (or Planck energy $E_p=hc/L_p)$.
%These two frameworks are traditionally regarded as physically distinct: in LIV models the relativity principle is explicitly broken, whereas in DSR it is deformed.

Much of the phenomenology of LIV and DSR is discussed at the level of one-particle dispersion relations of the form
\begin{equation}
E^2= {|\mathbf p|^2}+m^2+ f(E, {|\mathbf p|},m;E_p)
\label{eq1}
\end{equation}
where $m$ is the particle’s mass, $E$ and $ {|\mathbf p|}$ denote, respectively, its energy and spatial momentum and $c=1$. While a modified (non-quadratic) mass shell indeed signals a departure from the standard realization of Lorentz symmetry, it does not by itself uniquely determine the underlying relativistic structure. In particular, studies of $\kappa$-Poincar\'e kinematics have shown that nonlinear redefinitions of momentum variables can shift the deformation between the dispersion relation and the transformation laws \cite{MajidRuegg1994,KowalskiNowak2002}. Moreover, the relative-locality framework identifies the geometry of momentum space and the associated momentum-composition laws, rather than the one-particle dispersion relation alone, as the fundamental invariant structures \cite{RelativeLocality2011,Barcaroli2015}. Although these results suggest a nontrivial relation between Lorentz breaking and Lorentz deformation scenarios, a rigorous mathematical characterization of this relation, including its limitations, is still lacking.

Here, using Wigner's representation theory \cite{Wigner1939,WeinbergQFT} we prove that, under a global linearizability condition, the one-particle representation associated with a deformed mass shell is unitarily equivalent to the standard Wigner representation. As a consequence, Lorentz-violating and Lorentz-deformed kinematics correspond to different realizations of the same one-particle relativistic structure. For composite systems, the equivalence survives only when the momentum-composition law is affine-equivalent to ordinary momentum addition. The distinction between Lorentz breaking and Lorentz deformation is therefore encoded in genuine multiparticle observables carrying information about momentum composition laws. As an example, we discuss how Compton-edge measurements can distinguish between these two scenarios.

\section{One-particle equivalence and multiparticle obstruction}

We now show that, under a global linearizability condition, a modified one-particle mass provides a different realization of the same underlying one-particle relativistic structure, expressed in nonlinear momentum coordinates.
We consider the proper orthochronous Lorentz group $G=\mathrm{SO}^+(1, {d})$ and its universal cover $\widetilde G$, acting on one-particle momentum space. Let $V^*\simeq \mathbb{R}^{1, {d}}$ denote Minkowski momentum space and let $\Sigma_0^+(m)\subset V^*$ be the standard positive-energy Wigner orbit, i.e. the space of all on-shell relativistic momenta of a particle of mass $m$. We consider a smooth one-particle momentum manifold $\Sigma_f^+(m)$ corresponding to the experimentally selected positive-energy branch of a modified dispersion relation \eqref{eq1} equipped with a possibly nonlinear Lorentz action $\rho:\widetilde G\times \Sigma_f^+(m)\longrightarrow\Sigma_f^+(m)$, encompassing both LIV and DSR descriptions.

We call a one-particle mass shell globally linearizable if there exists 
a smooth, globally defined momentum remapping that commutes with the Lorentz action:  mathematically, a global equivariant diffeomorphism $ {F_m}:\Sigma_f^+(m)\longrightarrow\Sigma_0^+(m)$, such that
\begin{equation}
 {F_m}(\rho(\Lambda,p))=\Lambda\, {F_m}(p),
\label{eq:equiv-map}
\end{equation}
for all $\Lambda\in\widetilde G$. This condition states that the nonlinear Lorentz action on $\Sigma_f^+(m)$ is globally equivalent to the standard Wigner action under a change of momentum variables. Such maps are standard in the $\kappa$-Poincar\'e and relative-locality literature as nonlinear reparametrizations of momentum space, relating different coordinate descriptions of the same underlying momentum-space geometry. Here, however, $ {F_m}$ plays a different role: it tests whether the apparent deformation is intrinsic or merely reflects a nonlinear choice of momentum coordinates on the same relativistic orbit.

Under the assumption \eqref{eq:equiv-map}, the one-particle construction is completely determined. The invariant measure on $\Sigma_f^+(m)$ is obtained by change of 
variables through $F_m$ (pullback), $d\nu_{f,m}= {F_m^*}d\mu_{0,m}$,
where $d\mu_{0,m}$ is the standard Wigner measure on $\Sigma_0^+(m)$. The corresponding unitary intertwining operator is
\begin{equation}
(W_m\psi)(P)=\psi \left( {F_m^{-1}}(P)\right),
\label{eq:unitary-map}
\end{equation}
which identifies $L^2(\Sigma_f^+(m),d\nu_{f,m})$ with the standard Wigner Hilbert space. Pulling back the Poincar\'e action then gives
\begin{equation}
U_{f,m}(a,\Lambda)=W_m^{-1}U_{0,m}(a,\Lambda)W_m,
\label{eq:intertwining}
\end{equation}
where $U_{f,m}$ and $U_{0,m}$ denote the induced one-particle unitary representations associated with $\Sigma_f^+(m)$ and $\Sigma_0^+(m)$, respectively. The transported momenta are $Q^\mu=W_m^{-1} {\widehat P^\mu} W_m= {F_m^\mu}(p)$ , $Q^\mu Q_\mu=m^2$.

Thus, every globally linearizable one-particle kinematics is unitarily equivalent to the standard Wigner representation. A globally linearizable mass shell therefore carries no invariant distinction between Lorentz breaking and Lorentz deformation: the two correspond to different realizations of the same one-particle relativistic structure. The same modified dispersion relation may arise either from explicit Lorentz breaking in the variables $p_\mu$ or from a nonlinear realization of Lorentz symmetry in variables $ {F_m(p)}$, depending on the choice of momentum coordinates. Notice however that this equivalence is not merely a reparametrization of the dispersion relation, but a unitary equivalence of the full one-particle representation, including the action of boosts, spin degrees of freedom, invariant measures, and all associated representation-theoretic structures (see Appendix for details).

The theorem can be illustrated explicitly for the quartic dispersion relation $E^2- {|\mathbf p|^2}-\alpha( {|\mathbf p|^2})^2=m^2$, which provides a physically relevant benchmark. Quartic momentum corrections indeed arise as the leading even modification in the low-energy expansion of a broad class of Planck-scale dispersion relations (odd powers are removed if one assumes rotational invariance) \cite{Mattingly2005,Liberati2013}.
This example is also representative of Lorentz-deformed kinematics, since its one-particle dynamics is unitarily equivalent, up to $\mathcal{O}(E^3/E_P^3)$, to that of the DSR1 model \cite{MSF} and arises in a broad class of analogue gravity models \cite{NatRevPhys}. We therefore consider the positive-energy shell $\Sigma_\alpha^+={(E, {\mathbf p}) \in \mathbb{R}^{1+d}: E^2- {|\mathbf p|^2}-\alpha( {|\mathbf p|^2})^2=m^2,\  {E}>0}$, where $m>0$, $\alpha\ge0$, and $ {|\mathbf p|^2}=\delta^{ij}p_i p_j$. In the variables $(E, {\mathbf p})$ this shell appears Lorentz violating, since the standard linear Lorentz action does not preserve the quartic polynomial. Nevertheless, the radial map $ {F_\alpha}(E, {\mathbf p})=(E, {\mathbf f_\alpha}( {\mathbf p}))$ with $ {\mathbf f_\alpha}( {\mathbf p})=\sqrt{1+\alpha {|\mathbf p|^2}}\, {\mathbf p}$ defines a global shell-linearizing diffeomorphism. Indeed, $F_\alpha(\Sigma_\alpha^+))$ is exactly the ordinary positive-energy Wigner orbit, $\Pi_0^2-\Pi^2=m^2,\qquad \Pi_0>0$.
Since the radial derivative is positive for $\alpha\ge0$, the Jacobian $J_\alpha( {\mathbf p}) =(1+\alpha {|\mathbf p|^2})^{(d-2)/2}(1+2\alpha {|\mathbf p|^2})$ is everywhere positive, establishing that the construction is global rather than perturbative (see Appendix). Pulling back the invariant measure and the Wigner action therefore yields a unitary equivalence between the quartic and the standard one-particle representation.  The transported charges $ {Q^0}=E$ and $ {\mathbf Q}= {\mathbf f_\alpha}( {\mathbf p})$ obey $ {Q^\mu Q_\mu}=m^2$. At the level of single particles, the quartic shell thus provides no observable criterion for distinguishing Lorentz breaking from Lorentz deformation: both descriptions correspond to the same Wigner orbit, merely expressed in different momentum coordinates.

The ambiguity ends once particles are combined. The map $ {F_m}$ determines the geometry of a single mass shell, but it does not determine how momenta compose in multiparticle systems. The momentum-composition law is a kinematic ingredient that the one-particle dispersion relation does not determine  (equivalently, the coproduct of translations in Hopf-algebra language \cite{MajidBook}). A one-particle equivalence therefore extends to the two-particle translation sector only if the same map that linearizes the one-particle orbit also linearizes momentum composition, i.e. 
\begin{equation}
 {F}(p\oplus q)= {F}(p)+ {F}(q)
\label{eq:composition_condition}
\end{equation}
for all admissible pairs. The necessity of Eq.~(\ref{eq:composition_condition}) follows from the spectral theorem. In a two-particle state $|p, q\rangle$, the standard (primitive) translation by $a$ acts by the phase $e^{ia\cdot(F(p)+F(q))}$, while the DSR translation acts by $e^{ia\cdot F(p\oplus q)}$. For the map $W_m\otimes W_m$ to intertwine the two representations for all translation parameters $a$ and all momentum pairs, these phases must agree on every state, requiring $F(p\oplus q) = F(p)+F(q)$ to hold throughout the domain (see Appendix).

For ordinary fields, plane waves $ {e_p(X):=e^{ip\cdot X}}$ satisfy $e_p(X)e_q(X)=e_{p+q}(X)$, so the physical Fourier translation generators have the primitive coproduct $\Delta_{\rm phys}( {\widehat P_i})=  {\widehat P_i}\otimes  {\mathrm{Id}}+  {\mathrm{Id}}\otimes  {\widehat P_i}$.
If instead the auxiliary momentum coordinates $ {\mathbf f}(\mathbf p)$ are taken to be additive, with $ {F(E,\mathbf p)=(E,\mathbf f(\mathbf p))}$, and defining $ {\mathbf f^{-1}}=  {(\mathbf f)^{-1}}$ the induced DSR composition law becomes $\mathbf p \mathbin{ {\oplus_F}}\mathbf q =  {\mathbf f^{-1}} \left( {\mathbf f}(\mathbf p)+ {\mathbf f}(\mathbf q) \right)$, with coproduct $ {\Delta_F}( {\widehat P})= {\mathbf f^{-1}} \left( {\mathbf f}( {\widehat P})\otimes  {\mathrm{Id}}+  {\mathrm{Id}}\otimes {\mathbf f}( {\widehat P})\right)$. Different nonlinear extensions of the same one-particle shell map can therefore induce inequivalent momentum-composition laws.
For the quartic benchmark,
 $
\mathbf p\oplus_\alpha\mathbf q
=
\mathbf p+\mathbf q
-\frac{\alpha}{2}
\Big[
 {|\mathbf q|^2}\mathbf p
+ {|\mathbf p|^2}\mathbf q
+2(\mathbf p\cdot\mathbf q)
(\mathbf p+\mathbf q)
\Big]
+O(\alpha^2),
$ 
showing that the deformed total momentum differs from ordinary vector addition already at first order.

The obstruction is completely characterized by the following affine-rigidity theorem. Let $ {\mathbf f}:  {\mathbb R^d}\rightarrow  {\mathbb R^d}$ be a smooth diffeomorphism with $ {\mathbf f}(0)=0$ and inverse $ {\mathbf f^{-1}}$.
Then
\begin{equation}
\begin{aligned}
&( {W_m}\otimes  {W_m})\Delta_{\rm phys}( {\widehat P})( {W_m}\otimes  {W_m})^{-1}
= {\Delta_F}( {\widehat P})\
\qquad\Longleftrightarrow \ \\
& {\mathbf f}(\mathbf p+\mathbf q)=
 {\mathbf f}(\mathbf p)+ {\mathbf f}(\mathbf q)\quad  {\forall\,\mathbf p,\mathbf q\in\mathbb R^d}\
\qquad\Longleftrightarrow \\
& {\mathbf f}(\mathbf p)= {L}\mathbf p,
\end{aligned}
\label{eq:affine_rigidity}
\end{equation}
where $ {L}$ is a constant invertible matrix. Indeed, in the auxiliary momentum variables, the transported primitive coproduct has eigenvalues $ {\mathbf f^{-1}}(\mathbf\Pi)+ {\mathbf f^{-1}}(\mathbf\Upsilon)$,
whereas the DSR coproduct has eigenvalues $ {\mathbf f^{-1}}(\mathbf\Pi+\mathbf\Upsilon))$.
Equality of the total translation spectra for all momentum pairs therefore implies additivity of $ {\mathbf f^{-1}}$, and equivalently of $ {\mathbf f}$. Every smooth additive map is affine, so that $ {\mathbf f}(\mathbf p)= {L}\mathbf p$. Conversely, every affine map trivially intertwines the two coproducts (see Appendix). With the standard low-energy normalization $D {\mathbf f}(0)= {I_d}$, one further obtains $ {\mathbf f}=\mathrm{id}$.

Eq.~(\ref{eq:affine_rigidity}) identifies the precise point at which the one-particle equivalence fails to extend to composite systems. Since $ {\mathbf f_\alpha}(\mathbf p)$ is non-affine for every $\alpha>0$, the obstruction is nonperturbative and independent of the choice of boost basis. Bose or Fermi symmetrization can exchange particles, but it cannot alter the eigenvalue of the total translation generator, and noncommutative or nonassociative momentum-composition laws only strengthen the conclusion. We refer to this failure as the monoidal obstruction, borrowing the terminology from category theory, where a monoidal structure specifies how objects are combined \cite{MacLane1998}. Here it simply denotes the obstruction to lifting the one-particle equivalence to the composition law of composite systems. The one-particle mass shell can always be made to look ordinary by a nonlinear choice of momentum coordinates, but the momentum-composition law generally cannot. The invariant distinction between Lorentz-violating and Lorentz-deformed kinematics therefore resides not in the isolated one-particle shell, but in the genuine multiparticle structure encoded by momentum composition.

\section{Field algebra and reconstruction}

The same obstruction has a direct field-theoretic counterpart. In a field theory, the rule for multiplying fields already fixes the rule for composing momenta. This is already encoded in the above elementary plane-wave identity $ {e_p}(X) {e_q}(X)= {e_{p+q}}(X)$ which assigns additive Fourier momentum to the product. More generally, any (nonlinear) interaction term built from ordinary pointwise field multiplication — possibly dressed with translation-invariant operators such as derivatives, spin matrices, or nonlocal form 
factors — enforces ordinary additive momentum conservation. Such 
operators can modify the shape of an interaction vertex, but cannot 
alter the delta function imposed by the field product itself.

By the convolution theorem, every $n$-linear vertex has additive support (see Appendix)
\begin{equation}
\widehat{B}(p)=\int\prod_{a=1}^{n}[\mathrm{d}p_a]
V_B(p;p_1,\dots,p_n) (2\pi)^D \delta^{(D)}\!\Big(p-\sum_{a=1}^{n}p_a\Big) \prod_{a=1}^{n}\widehat{\psi}_a(p_a) 
\label{eq:additive-support}
\end{equation}
where $ {D=d+1}$, $p=(p^0,\mathbf p)$. All model dependence is contained in the vertex tensor $V_B$, while the delta function enforcing additive momentum conservation is fixed by the ordinary field product. Consequently, higher-derivative operators, matrix-valued kinetic terms, and translation-invariant nonlocal kernels can modify the one-particle spectrum and interaction vertices, but not the additive conservation law. A modified dispersion relation realized within an ordinary field algebra therefore represents only a one-particle spectral modification, not by itself a deformed relativity principle.

Conversely, a DSR translation sector is obtained only by reconstructing the field algebra together with the translation generators. If plane waves satisfy $ {e_p}\mathbin{ {\star_F}} {e_q}=e_{ {F^{-1}}( {F}(p)+ {F}(q))}$ with $D_i  {e_p}=i {F_i}(p) {e_p}$, then the operators $D_i$ are derivations of the $ {\star_F}$-product and interaction vertices conserve the additive auxiliary momentum $ {F}(p)$. In the Abelian setting considered here the product remains associative and commutative, but becomes nonlocal in the original coordinates whenever $ {F}$ is nonlinear. For the quartic benchmark,$D_i=(1-\alpha\Delta)^{1/2} \partial_i$, so that translations are generated by pseudodifferential rather than finite-order differential operators. The DSR sector is therefore not a rewriting of the ordinary local theory: it is obtained only by reconstructing, in a mutually compatible way, both the field product and the translation generators (see Appendix).

\section{Experimental tests of momentum-composition law}

Once the one-particle shell has been reconstructed, every observable depending solely on the dispersion relation is identical in the two descriptions. Observable differences therefore arise only in genuinely multiparticle processes that are sensitive to the total translation character of composite states, or equivalently, to the momentum-composition law. Representative examples include pair production thresholds \cite{Aharonian2006Threshold} and recoil measurements \cite{Morel2020Recoil}, invariant-mass reconstruction \cite{Lees2017MissingMass}, and interferometric observables \cite{Bouchendira2011RamseyBorde}, among many others \cite{Burke1997BreitWheeler,Abramowicz2015DIS,Achard2005Bhabha,Aubert2009RadiativeReturn,Akushevich2000Triplet,Bethe1934PairProduction,AguilarArevalo2015Pion,Aker2019Beta}. In each case, the relevant observable is fixed by a conservation law involving several external momenta and can therefore retain information about how momenta are combined.

\begin{figure}[!t]
\centering
\includegraphics[width=0.995\columnwidth]{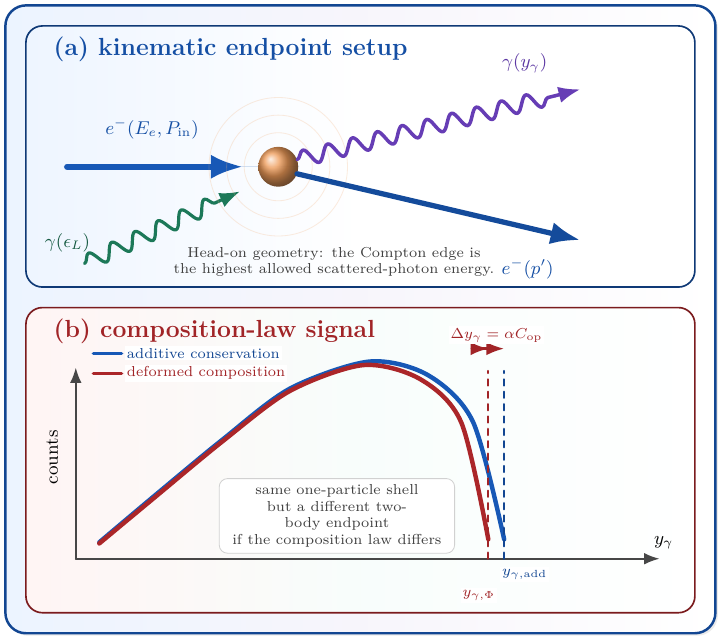}
\caption{(a) Head-on collision between 
an electron of laboratory energy $E_e$ and spatial momentum $P_{\rm in}$ ad a photon of energy $\epsilon_L$, producing an outgoing photon of energy $y_\gamma$ and a recoiling electron with momentum $p'$. (b) Photon-energy spectra obtained with ordinary additive momentum conservation (blue) and with a deformed momentum-composition law (red). The endpoint displacement isolates the composition-law contribution after the common one-particle shell has been calibrated away.}
\label{fig:comptonedge}
\end{figure}

As a concrete example, we consider Compton-edge measurements \cite{Bocquet2010Compton,Gharibyan2012}, where the observable is the endpoint of the energy spectrum of photons backscattered in the head-on collision of a relativistic electron with a laser photon. The precise determination of this endpoint is routinely used in Compton polarimetry for energy calibration and beam diagnostics and has also been used to constrain additive LIV \cite{Bocquet2010Compton}. Because the endpoint is fixed by energy-momentum conservation, it provides a direct probe of the momentum-composition law. The idea is schematically illustrated in Fig. \ref{fig:comptonedge}. Since LIV and DSR have identical one-particle shells but different multiparticle conservation laws, once the common shell deformation and detector response are fixed independently, the residual endpoint separation directly isolates the composition-law contribution.

Let $\alpha_{\rm sh}$ denote the common quartic shell coefficient and let $\mathcal C_{\rm op}$ denote the Compton-edge response coefficient fixed by the experimental apparatus (beam energy,
laser frequency, collision geometry, detector response, etc.). Primitive additive conservation retains the leading shell-induced endpoint shift, $-\alpha_{\rm sh}\mathcal C_{\rm op}$, whereas the map-induced DSR composition law provides a compensating contribution. Thus, if $y$ denotes the Compton-edge position, one finds to first order (see Appendix)
\begin{equation}
\left|y_{\rm LIV}-y_{\rm DSR}^{\rm match}\right|=\left|y_{\rm LIV}-y_{\rm SR}\right|\left|\alpha_{\rm sh}\mathcal C_{\rm op}\right|+\mathcal O(\alpha_{\rm sh}^{2}) .
\label{eq:compton-residual}
\end{equation}
Therefore, at first order the LIV-DSR separation has exactly the same magnitude as the LIV-SR separation. As said, this equality follows from the DSR composition-law correction compensating the shell-induced shift, while no such compensation occurs for additive LIV.

Current astrophysical observations already constrain the scale of quartic departures from Lorentz symmetry. Taking the $95\%$ confidence-level lower limit of $1.14\,\mathrm{PeV}$ on the photon cutoff reported by the LHAASO analysis of J2032+4102 \cite{Cao2022LHAASO} the positive common photon--electron quartic model considered here gives
$\alpha<8.25\times10^{-31}\,\mathrm{GeV}^{-2}$,
or equivalently to a lower bound
$\Lambda_\alpha>1.10\times10^{15}\,\mathrm{GeV}$
on the scale suppressing quartic corrections. The same one-particle coefficient enters both the additive-LIV description and its matched DSR realization. Equation~(\ref{eq:compton-residual}) therefore translates the astrophysical bound into a maximum composition-induced endpoint displacement within the benchmark model. For an $11,\mathrm{GeV}$ electron beam and a $532,\mathrm{nm}$ laser we obtain $\lvert\Delta y_{\rm LIV-SR}\rvert=\lvert\Delta y_{\rm LIV-DSR}^{\rm match}\rvert<1.59\times10^{-10}\,\mathrm{eV}$, corresponding to $\lvert\Delta y\rvert/y_0<5.13\times10^{-20}$ (see Appendix for a detailed discussion). While such an endpoint shift lies well below the sensitivity achieved in existing Compton-edge measurements \cite{Bocquet2010Compton,Gharibyan2012}, future higher-energy facilities could nevertheless substantially improve the experimental bounds, since the kinematic coefficient $\mathcal C_{\rm op}$ grows rapidly with beam energy.  Moreover, a measurement performed at several beam energies and laser frequencies would test the characteristic kinematic scaling of $\mathcal C_{\rm op}$, providing a way to distinguish a composition-law residual from energy-independent instrumental offsets and other systematic effects.

\section{Conclusions} 

We have shown that, under the assumption of global shell linearizability, Lorentz-violating and Lorentz-deformed kinematics are physically equivalent at the one-particle level. The two descriptions correspond to unitarily equivalent realizations of the same Wigner orbit expressed in different momentum coordinates. Consequently, a modified one-particle dispersion relation does not by itself define a deformed relativity principle. The distinction emerges only when one specifies how momenta compose. We proved that the one-particle equivalence extends to composite systems if and only if the momentum map is affine, so that the deformed composition law reduces to ordinary addition. The quartic model concretely illustrates this general result: its modified shell is globally linearizable, yet its nonlinear momentum-composition law cannot be reduced to ordinary addition because of affine rigidity. 

Our results imply that, under the conditions of our theorem, single-particle observables cannot distinguish Lorentz violation from Lorentz deformation, even though they may provide signatures of Planck-scale physics. The distinction resides instead in the momentum-composition law. We discussed how Compton-edge measurements could potentially isolate the composition-law contribution to energy-momentum conservation after the common one-particle shell has been calibrated away. For the quartic benchmark, the latter provides a concrete estimate of the sensitivity to the energy scale suppressing non-affine momentum composition. The search for the underlying structure of quantum spacetime should therefore move beyond isolated-particle spectroscopy toward a tomography of composition laws, by asking how particles compose and how conservation laws act on multiparticle states.

\section{Appendix}
\appendix

\subsection{One-particle equivalence: the transported representation and the quartic benchmark}

We prove that (2)-(4) define a unitary equivalence of the specified transported one-particle representations and verify that the quartic shell is globally linearizable. We use $c=\hbar=1$, $D=d+1$, and $\eta=\operatorname{diag}(+,-,\ldots,-)$.

\paragraph{Invariant measure and unitarity of \texorpdfstring{$W_m$}{Wm}.}

The one-particle momentum manifold $\Sigma_f^+(m)$ is the selected connected, regular positive-energy branch of
\begin{equation}
 C_f(p;m)=E^2-|\mathbf p|^2-m^2-f(E,|\mathbf p|,m;E_p)=0.
 \label{s1:shell}
\end{equation}
It carries a smooth $\Ge$-action $\rho$ as assumed in Eq.~(2); the shell alone does not specify this action. Given a global equivariant diffeomorphism $F_m$ satisfying  Eq.~(2), define the pullback measure
\begin{equation}
 \dd\nu_{f,m}=F_m^*\dd\mu_{0,m},\label{s1:measure}
\end{equation}
where $\dd\mu_{0,m}(P)=\dd^d\mathbf P/(2P^0)$ is the standard Lorentz-invariant measure on the Wigner orbit \cite{Wigner1939,WeinbergQFT}. Equivariance of $F_m$ makes $\dd\nu_{f,m}$ invariant under $\rho$:
\begin{equation}
 \rho(\Lambda,\cdot)^*\dd\nu_{f,m}
 =(F_m\circ\rho(\Lambda,\cdot))^*\dd\mu_{0,m}
 =(\Lambda\circ F_m)^*\dd\mu_{0,m}
 =F_m^*\dd\mu_{0,m}=\dd\nu_{f,m}.\label{s1:invariance}
\end{equation}
The intertwiner $(W_m\psi)(P)=\psi(F_m^{-1}(P))$ maps $L^2(\Sigma_f^+(m),\dd\nu_{f,m})$ onto $L^2(\Sigma_0^+(m),\dd\mu_{0,m})$ and is unitary by the change of variables $P=F_m(p)$:
\begin{equation}
 \|W_m\psi\|^2=\int_{\Sigma_0^+(m)}|\psi(F_m^{-1}(P))|^2\dd\mu_{0,m}(P)
 =\int_{\Sigma_f^+(m)}|\psi(p)|^2\dd\nu_{f,m}(p)=\|\psi\|^2.\label{s1:unitarity}
\end{equation}
The inverse is $(W_m^{-1}\phi)(p)=\phi(F_m(p))$. Matrix elements, orthogonality relations, and spectral probabilities of correspondingly transported states and observables are therefore identical in the two representations.
In a spatial graph chart $\iota_f(\mathbf k)=(E_f(\mathbf k),\mathbf k)$ the pullback measure is explicitly
\begin{equation}
 \dd\nu_{f,m}=\frac{|\det D_{\mathbf k}\mathbf P(\mathbf k)|}{2\sqrt{m^2+|\mathbf P(\mathbf k)|^2}}\dd^d\mathbf k,
 \qquad \mathbf P(\mathbf k)=\operatorname{pr}_{\rm sp}F_m(\iota_f(\mathbf k)).\label{s1:chart}
\end{equation}
No energy-preserving form of a general $F_m$ is assumed here. For the quartic benchmark below, $F_\alpha(E,\mathbf p)=(E,\mathbf f_\alpha(\mathbf p))$.

\paragraph{Transported representation and spin.}
The transported representation of Eq. 4 inherits strong continuity and the group law from $U_{0,m}$ by conjugation. The Wigner mass $m$ and spin or helicity sector are fixed; the shell alone does not determine the spin or helicity data. For a scalar particle its action is
\begin{equation}
 (U_{f,m}(a,\Lambda)\psi)(p)=e^{ia\cdot F_m(p)}\psi\bigl(\rho(\Lambda^{-1},p)\bigr),\label{s1:scalar}
\end{equation}
so the translation generators are transported to $Q^\mu=F_m^\mu(p)$, satisfying $Q^\mu Q_\mu=m^2$, and the Casimir eigenvalues are unchanged. For $d=3$, $m>0$, and spin $s$, the Wigner representation acts on $L^2(\Sigma_0^+(m),\dd\mu_{0,m})\otimes\mathbb C^{2s+1}$ by
\begin{revblock}
\begin{equation}
 \begin{split}
 [U_{0,m}(a,\Lambda)\Phi]_\sigma(P)
 &=e^{ia\cdot P}\sum_{\tau=-s}^{s}D^{(s)}_{\sigma\tau}\bigl(R(\Lambda,P)\bigr)\Phi_\tau(\Lambda^{-1}P),\\
 R(\Lambda,P)&=B(P)^{-1}\Lambda B(\Lambda^{-1}P).
 \end{split}\label{s1:wigner}
\end{equation}
\end{revblock}
Here $B(P)\in\Ge$ is a standard boost to $P$ and $R(\Lambda,P)$ is the Wigner rotation in $SU(2)$. Since $W_m$ acts componentwise, $(W_m\Psi)_\sigma(P)=\Psi_\sigma(F_m^{-1}(P))$, conjugation replaces $P$ by $F_m(p)$ in Eq.~\eqref{s1:wigner}:
\begin{equation}
 [U_{f,m}(a,\Lambda)\Psi]_\sigma(p)
 =e^{ia\cdot F_m(p)}\sum_\tau D^{(s)}_{\sigma\tau}\bigl(R(\Lambda,F_m(p))\bigr)
 \Psi_\tau\bigl(\rho(\Lambda^{-1},p)\bigr).\label{s1:spin}
\end{equation}
The little-group structure, spin multiplicity, and Pauli-Lubanski eigenvalue $w_f^\mu w_{f\mu}=-m^2s(s+1)\Id$ are all unchanged.

For the photon, use the punctured forward cone $\Sigma_0^+(0)=\{P:P^2=0,P^0>0\}$ and measure $\dd^3\mathbf P/(2|\mathbf P|)$. In each finite-helicity sector the $SU(2)$ matrix is replaced by the helicity phase $e^{ih\theta(\Lambda,P)}$; the translational part of the massless little group acts trivially. The same pullback gives $Q^2=0$ and $w_f^\mu=hQ^\mu$, with the cone apex excluded.

For a densely defined closed observable $A_0$, $A_f=W_m^{-1}A_0W_m$ has domain $W_m^{-1}\operatorname{Dom}(A_0)$ and adjoint $A_f^*=W_m^{-1}A_0^*W_m$. For self-adjoint $A_0$, its spectral projections and Borel functional calculus are conjugated by the same unitary. These statements justify the full transported-observable claim; they do not identify the Fourier momentum $\widehat p^\mu$ with the transported charge $Q^\mu=F_m^\mu(\widehat p)$. Thus equality of probabilities requires the states and observables to be transported together. It does not follow from the common shell alone that an independently specified LIV theory has these translations, boosts or measurement couplings.

\paragraph{Global linearizability of the quartic benchmark.}
For the quartic shell $\Sigma_\alpha^+(m)=\{E^2-|\mathbf p|^2-\alpha|\mathbf p|^4=m^2,E>0\}$, the map $F_\alpha(E,\mathbf p)=(E,\mathbf f_\alpha(\mathbf p))$ with
\begin{equation}
 \mathbf f_\alpha(\mathbf p)=\sqrt{1+\alpha|\mathbf p|^2}\,\mathbf p\label{s1:map}
\end{equation}
satisfies the ambient identity
\begin{revblock}
\begin{equation}
 C_0(F_\alpha(p);m)=E^2-|\mathbf f_\alpha(\mathbf p)|^2-m^2
 =E^2-|\mathbf p|^2-\alpha|\mathbf p|^4-m^2=C_\alpha(p;m),\label{s1:shellidentity}
\end{equation}
\end{revblock}
so it maps every quartic shell onto the corresponding Wigner orbit. The radial function $g(r)=r\sqrt{1+\alpha r^2}$ satisfies $g'(r)=(1+2\alpha r^2)/\sqrt{1+\alpha r^2}>0$ for all $r\ge0$ and all $\alpha\ge0$, with $g(0)=0$ and $g(r)\to\infty$, so $\mathbf f_\alpha$ is a global diffeomorphism of $\mathbb R^d$. Its inverse is
\begin{equation}
 \mathbf f_\alpha^{-1}(\bm\Pi)=\left(\frac{2}{1+\sqrt{1+4\alpha|\bm\Pi|^2}}\right)^{1/2}\bm\Pi,\label{s1:inverse}
\end{equation}
which is smooth on all of $\mathbb R^d$ (the prefactor is analytic in $|\bm\Pi|^2$). The Jacobian is
\begin{equation}
 J_\alpha(\mathbf p)=\det D\mathbf f_\alpha(\mathbf p)
 =(1+\alpha|\mathbf p|^2)^{(d-2)/2}(1+2\alpha|\mathbf p|^2)>0,\label{s1:jac}
\end{equation}
positive everywhere, so the pullback measure $\dd\nu_{\alpha,m}=J_\alpha(\mathbf p)\dd^d\mathbf p/(2E_{\alpha,m}(\mathbf p))$ is smooth and globally defined on the selected shell (for $m=0$, away from the apex). Equation~\eqref{s1:unitarity} then gives unitarity of $W_m$ for all $m\ge0$ and $\alpha\ge0$ without any perturbative expansion in $\alpha$.
The transported boost vector fields are, for an infinitesimal rapidity $\delta\eta$ along direction $i$,
\begin{equation}
 X_i=f_{\alpha,i}(\mathbf p)\partial_E+E\sum_j(D\mathbf f_\alpha)^{-1}_{ji}\partial_{p_j},\label{s1:boost}
\end{equation}
where, with $r=|\mathbf p|$ and Euclidean spatial components,
\begin{revblock}
\begin{equation}
 (D\mathbf f_\alpha)^{-1}_{ji}=\frac{1}{\sqrt{1+\alpha r^2}}
 \left(\delta_{ji}-\frac{\alpha p_jp_i}{1+2\alpha r^2}\right).\label{s1:derivative}
\end{equation}
\end{revblock}
Differentiating Eq.~\eqref{s1:shellidentity} gives $X_iC_\alpha=2Ef_{\alpha,i}-2E\sum_{j,k}f_{\alpha,k}(D\mathbf f_\alpha)_{kj}(D\mathbf f_\alpha)^{-1}_{ji}=0$. Also $X_i\to p_i\partial_E+E\partial_{p_i}$ as $\alpha\to0$. The finite flow of $X_i$ is $F_\alpha^{-1}\circ\Lambda\circ F_\alpha$, defined globally on every positive-energy branch because $F_\alpha$ is a global diffeomorphism.

For the truncated effective relation $E_\psi^2=|\mathbf p|^2(1+E_\psi/E_p)+m^2$, displayed approximately in Eq.~(1) of Ref.~[29], completing the square with $E_\varphi=E_\psi-|\mathbf p|^2/(2E_p)$ gives $E_\varphi^2=m^2+|\mathbf p|^2+|\mathbf p|^4/(4E_p^2)$. This algebraic identity is exact for the truncated relation; the quartic coefficient is not a controlled second-order prediction for the untruncated model and does not specify its momentum-composition law.

\subsection{Momentum composition, affine rigidity and the reconstruction of the field algebra}
\label{sec:s2}
We derive Eqs.~(5)-(7), the quartic composition law, and the compatible field product and derivations used in the main text.

The shell map $F_m$ is defined on $\Sigma_f^+(m)$ only, whereas the sum of two one-particle momenta is in general off shell. To define the map-induced composition law, choose an ambient diffeomorphism $F$ of $\mathbb R^{1,d}$ with $F(0)=0$, the same for all species, whose restrictions to the various shells are the shell maps. Given such an $F$, the map-induced law and its inverse are~\cite{JudesVisser2003,RelativeLocality2011}
\begin{equation}
 p\oplus_Fq=F^{-1}\bigl(F(p)+F(q)\bigr),\qquad
 \ominus_Fp=F^{-1}\bigl(-F(p)\bigr).
 \label{s2:composition}
\end{equation}
Applying $F$ to either bracketing of a three-particle total gives $F(p)+F(q)+F(r)$, so the law is associative and commutative, with neutral element $0$; and $\Lambda_F=F^{-1}\Lambda F$ satisfies $\Lambda_F(p\oplus_Fq)=\Lambda_Fp\oplus_F\Lambda_Fq$. Eq.~(5), $F(p\oplus q)=F(p)+F(q)$, is satisfied identically by this law: it is the statement that the auxiliary charges $F(p)$ add. For the form $F(E,\mathbf p)=(E,\mathbf f(\mathbf p))$, the energies add and the spatial law is $\mathbf p\oplus_F\mathbf q=\mathbf f^{-1}(\mathbf f(\mathbf p)+\mathbf f(\mathbf q))$, the induced DSR composition law of the main text.

A fixed-mass shell map does not specify its ambient extension, so composition is additional kinematical data. For the quartic benchmark, the global, species-independent ambient map $F_\alpha$ of Eq.~\eqref{s1:map} defines the law for all pairs, with
\begin{equation}
 \mathbf p\oplus_\alpha\mathbf q
 =\mathbf f_\alpha^{-1}\bigl(\mathbf f_\alpha(\mathbf p)+\mathbf f_\alpha(\mathbf q)\bigr),\qquad
 E\oplus_\alpha E'=E+E'.
 \label{s2:quarticlaw}
\end{equation}
For a near-identity map $F_\varepsilon(p)=p+\varepsilon U(p)+O(\varepsilon^2)$ with $U(0)=0$, inversion gives
\begin{equation}
 p\oplus_{F_\varepsilon}q=p+q+\varepsilon\bigl[U(p)+U(q)-U(p+q)\bigr]+O(\varepsilon^2),
 \label{s2:cauchydefect}
\end{equation}
so the leading deviation from ordinary addition is the Cauchy defect of $U$. For the quartic map $U(\mathbf p)=\tfrac12|\mathbf p|^2\mathbf p$ by expansion of Eq.~\eqref{s1:map}, and the defect is
\begin{equation}
 \begin{split}
 \mathbf p\oplus_\alpha\mathbf q&=\mathbf p+\mathbf q+\alpha\bm\Xi^{(1)}(\mathbf p,\mathbf q)+O(\alpha^2),\\
 \bm\Xi^{(1)}(\mathbf p,\mathbf q)&=-\frac12\bigl[|\mathbf q|^2\mathbf p+|\mathbf p|^2\mathbf q
 +2(\mathbf p\cdot\mathbf q)(\mathbf p+\mathbf q)\bigr],
 \end{split}
 \label{s2:defect}
\end{equation}
which is the first-order law discussed in the main text. The expansions khere are
\begin{align}
 \mathbf f_\alpha(\mathbf p)&=\mathbf p+\frac\alpha2|\mathbf p|^2\mathbf p+O(\alpha^2),\\
 \mathbf f_\alpha^{-1}(\bm\Pi)&=\bm\Pi-\frac\alpha2|\bm\Pi|^2\bm\Pi+O(\alpha^2),
\end{align}
at fixed momenta as $\alpha\to0$.
For collinear momenta $a\mathbf n$ and $b\mathbf n$ the leading defect is
\begin{equation}
 \bm\Xi^{(1)}(a\mathbf n,b\mathbf n)=-\tfrac32 ab(a+b)\mathbf n,
 \label{s2:collinear}
\end{equation}
which vanishes only when one momentum vanishes or $b=-a$. The first-order expansion \eqref{s2:defect} approximates the exact law \eqref{s2:quarticlaw}, whose associativity follows from its definition as the pullback of ordinary addition.

We now turn to the two-particle translation sector and to the proof of Eq.~(6). Let $H_m\otimes H_{m'}$ be the two-particle space with the product measure, and transform both factors to auxiliary coordinates by $W_m\otimes W_{m'}$, so that wave functions become functions $\Phi(\bm\Pi,\bm\Upsilon)$ of the auxiliary spatial momenta (the energies are determined by the Wigner shells). In Eq.~(6), $\widehat{\mathbf P}$ denotes the physical Fourier momentum, multiplication by the spatial momentum $\mathbf p$ in the physical coordinates; in the auxiliary coordinates it is multiplication by $\mathbf f^{-1}(\bm\Pi)$. The primitive coproduct $\Delta_{\rm phys}(\widehat{\mathbf P})=\widehat{\mathbf P}\otimes\Id+\Id\otimes\widehat{\mathbf P}$ and the map-induced coproduct $\Delta_F(\widehat{\mathbf P})=\mathbf f^{-1}(\mathbf f(\widehat{\mathbf P})\otimes\Id+\Id\otimes\mathbf f(\widehat{\mathbf P}))$ are therefore the multiplication operators by $\mathbf f^{-1}(\bm\Pi)+\mathbf f^{-1}(\bm\Upsilon)$ and by $\mathbf f^{-1}(\bm\Pi+\bm\Upsilon)$ respectively, and the two translation groups they generate act as
\begin{equation}
 \begin{split}
 [T_{\rm phys}(\mathbf a)\Phi](\bm\Pi,\bm\Upsilon)
 &=e^{i\mathbf a\cdot[\mathbf f^{-1}(\bm\Pi)+\mathbf f^{-1}(\bm\Upsilon)]}\Phi(\bm\Pi,\bm\Upsilon),\\
 [T_F(\mathbf a)\Phi](\bm\Pi,\bm\Upsilon)
 &=e^{i\mathbf a\cdot\mathbf f^{-1}(\bm\Pi+\bm\Upsilon)}\Phi(\bm\Pi,\bm\Upsilon).
 \end{split}
 \label{s2:translations}
\end{equation}
Since $F(E,\mathbf p)=(E,\mathbf f(\mathbf p))$ leaves the energy untouched, the time-translation parts of the two coproducts coincide and only the spatial parts need comparison. These are the totals of the Fourier labels $\mathbf p$, compared under the fixed product identification. They are distinct from the additive transported charge $\mathbf f(\mathbf p)+\mathbf f(\mathbf q)$. Their use as a comparison of measured momenta requires a common momentum calibration; this is additional physical input. Each real multiplication operator is understood on its maximal domain $\{\Phi\in L^2:h\Phi\in L^2\}$, where $h$ is its multiplier; equality of operators includes equality of these domains.

\Needspace{10\baselineskip}
\begin{theorem}[affine rigidity, Eq.~(6)]\label{s2:rigidity}
Let $\mathbf f:\mathbb R^d\to\mathbb R^d$ be a smooth diffeomorphism with $\mathbf f(0)=0$. The following are equivalent:
\begin{enumerate}
\item[(a)] $\Delta_{\rm phys}(\widehat{\mathbf P})=\Delta_F(\widehat{\mathbf P})$, equivalently $T_{\rm phys}(\mathbf a)=T_F(\mathbf a)$ for all $\mathbf a\in\mathbb R^d$;
\item[(b)] $\mathbf f(\mathbf p+\mathbf q)=\mathbf f(\mathbf p)+\mathbf f(\mathbf q)$ for all $\mathbf p,\mathbf q\in\mathbb R^d$;
\item[(c)] $\mathbf f(\mathbf p)=L\mathbf p$ for a constant invertible matrix $L$.
\end{enumerate}
If in addition $D\mathbf f(0)=\Id$, then $\mathbf f=\mathrm{id}$.
\end{theorem}
\begin{proof}
(a)$\Rightarrow$(b). By Stone's theorem, a strongly continuous unitary one-parameter group determines its self-adjoint generator; equality of $T_{\rm phys}(t\mathbf a)$ and $T_F(t\mathbf a)$ for all $t$ and all directions $\mathbf a$ gives equality of the generators, i.e. of the two multiplication operators in Eq.~\eqref{s2:translations}. Two multiplication operators on $L^2$ of a product measure coincide if and only if their multipliers agree almost everywhere. The product measure has full support on $\mathbb R^d\times\mathbb R^d$ (the spatial projection of $\Sigma_0^+(m)$ is $\mathbb R^d$, with the origin removed for $m=0$, and $\dd^d\bm\Pi/(2P^0)$ has a positive density there), and both multipliers are continuous; two continuous functions that agree almost everywhere with respect to a measure of full support agree everywhere. Hence $\mathbf f^{-1}(\bm\Pi+\bm\Upsilon)=\mathbf f^{-1}(\bm\Pi)+\mathbf f^{-1}(\bm\Upsilon)$ for all $\bm\Pi,\bm\Upsilon$, which is the additivity of $\mathbf f^{-1}$; writing $\mathbf p=\mathbf f^{-1}(\bm\Pi)$, $\mathbf q=\mathbf f^{-1}(\bm\Upsilon)$ and applying $\mathbf f$ gives (b). This proves the necessity of Eq.~(5) for primitive addition under the specified product identification.

(b)$\Rightarrow$(c). Differentiating $\mathbf f(\mathbf p+\mathbf q)=\mathbf f(\mathbf p)+\mathbf f(\mathbf q)$ with respect to $\mathbf q$ at $\mathbf q=0$ gives $D\mathbf f(\mathbf p)=D\mathbf f(0)=:L$ for every $\mathbf p$. Integrating along the segment from $0$ to $\mathbf p$, $\mathbf f(\mathbf p)=\mathbf f(0)+\int_0^1D\mathbf f(t\mathbf p)\mathbf p\,\dd t=L\mathbf p$; $L$ is invertible because $\mathbf f$ is a diffeomorphism.

(c)$\Rightarrow$(a). If $\mathbf f(\mathbf p)=L\mathbf p$, then $\mathbf f^{-1}=L^{-1}$ is linear and $\mathbf f^{-1}(\bm\Pi+\bm\Upsilon)=\mathbf f^{-1}(\bm\Pi)+\mathbf f^{-1}(\bm\Upsilon)$ identically, so the two lines of Eq.~\eqref{s2:translations} coincide. This is the statement of the main text that every affine map fixing the origin trivially intertwines the two coproducts. Finally, $D\mathbf f(0)=\Id$ forces $L=\Id$.
\end{proof}

The theorem compares totals under the fixed $W_m\otimes W_{m'}$ identification, not under arbitrary unitaries. Since $\mathbf f_\alpha$ is nonaffine for every $\alpha>0$, the two totals differ on an open set of momentum pairs, nonperturbatively. The map-induced law itself is linearized by $F$, as Eq.~\eqref{s2:composition} shows; the obstruction concerns agreement with primitive addition of the same Fourier momenta.

Both multipliers in Eq.~\eqref{s2:translations} are symmetric under the exchange $\bm\Pi\leftrightarrow\bm\Upsilon$, so for identical particles the symmetrizer and the antisymmetrizer commute with both translation groups, and the theorem restricts unchanged to the bosonic and fermionic subspaces. Explicitly, if the defect $\mathbf f^{-1}(\bm\Pi+\bm\Upsilon)-\mathbf f^{-1}(\bm\Pi)-\mathbf f^{-1}(\bm\Upsilon)$ is nonzero at $(\bm\Pi_*,\bm\Upsilon_*)$, it is nonzero on a neighborhood containing a pair with distinct momenta. Symmetric or antisymmetric packets $g(\bm\Pi)h(\bm\Upsilon)\pm g(\bm\Upsilon)h(\bm\Pi)$ with $g,h$ supported in small disjoint balls around that pair are supported where the defect is nonzero; on such packets $T_{\rm phys}(\mathbf a)\ne T_F(\mathbf a)$ for suitable $\mathbf a$. Symmetrization exchanges particles but cannot alter the eigenvalue of the total translation generator, as stated in the main text.

A noncommutative or nonassociative composition law cannot obey Eq.~(5) for any injective $F$, since that identity would make the law inherit commutativity and associativity from addition. The bicrossproduct $\kappa$-Poincar\'e coproduct is an example of a noncommutative, coassociative coproduct~\cite{MajidBook,MajidRuegg1994}. For $n$ particles the map-induced total is $F^{-1}(\sum_{a=1}^nF(p_a))$ while the ordinary field product supplies $\sum_{a=1}^np_a$; for a nonaffine map these disagree on some two-particle configurations, preventing agreement of the full composition prescriptions. This failure to identify primitive and map-induced totals under the fixed one-particle identification is the composition obstruction referred to in the main text

The obstruction has a direct field-theoretic counterpart, Eq.~(7), which follows from the convolution theorem. Fields are written as $\psi(X)=\int[\dd p]\,\widehat\psi(p)e_p(X)$ with $e_p(X)=e^{ip\cdot X}$, $p\cdot X=p^0X^0-\mathbf p\cdot\mathbf X$, and $[\dd p]=\dd^Dp/(2\pi)^D$ as in Eq.~(7). Pointwise multiplication of plane waves gives $e_pe_q=e_{p+q}$, and for two fields the convolution theorem reads
\begin{equation}
 \widehat{\psi\phi}(p)=\int[\dd p_1][\dd p_2]\,(2\pi)^D\delta^{(D)}(p-p_1-p_2)\widehat\psi(p_1)\widehat\phi(p_2).
 \label{s2:convolution}
\end{equation}
A translation-invariant Fourier multiplier $K$ acts by $(K\psi)^\wedge(p)=K(p)\widehat\psi(p)$; derivatives ($K(p)=ip_\mu$), spin matrices, internal contractions and nonlocal form factors are all of this type. The constructions are understood on a common test domain on which the indicated products and Fourier multipliers are defined.

\begin{proposition}[Eq.~(7)]\label{s2:supportprop}
Every $n$-linear vertex built from pointwise products and translation-invariant Fourier multipliers has the form
\begin{equation}
 \widehat B(p)=\int\prod_{a=1}^n[\dd p_a]\,V_B(p;p_1,\ldots,p_n)(2\pi)^D
 \delta^{(D)}\!\left(p-\sum_{a=1}^np_a\right)\prod_{a=1}^n\widehat\psi_a(p_a),
 \label{s2:addsupport}
\end{equation}
with all model dependence in the vertex tensor $V_B$.
\end{proposition}
\begin{proof}
By induction on the construction. A single field is of the form \eqref{s2:addsupport} with $n=1$ and $V_B=1$. Applying a multiplier $K$ to an expression of the form \eqref{s2:addsupport} multiplies $V_B$ by $K(p)$; applying it to a subexpression containing the fields indexed by $A$ multiplies $V_B$ by $K(\sum_{a\in A}p_a)$. Multiplying two expressions of the form \eqref{s2:addsupport}, with $n_1$ and $n_2$ fields, by Eq.~\eqref{s2:convolution} produces one more momentum integration and the product of the two delta functions with $\delta^{(D)}(p-p'-p'')$, which integrates to a single delta function enforcing $p=\sum_{a=1}^{n_1+n_2}p_a$. In each step the delta function keeps its additive argument and only $V_B$ changes.
\end{proof}
Consequently, within this class of constructions, higher-derivative kinetic terms, matrix-valued operators and translation-invariant nonlocal kernels modify the one-particle spectrum and the vertex tensors but never the additive conservation law. A modified dispersion relation realized inside an ordinary field algebra has the primitive coproduct $\Delta_{\rm phys}$; with the quartic shell this is precisely the additive Lorentz-violating realization compared in Sec.~S3.

Conversely, the map-induced DSR translation sector is obtained by reconstructing the field algebra together with its translation generators. Let $F$ be a global ambient diffeomorphism as in Eq.~\eqref{s2:composition}. The following formulas define an algebra on fields with smooth compactly supported Fourier transforms. For the quartic map, $F_\alpha$ and $F_\alpha^{-1}$ and their derivatives have polynomial bounds, the radial maps grow at positive powers at infinity, and $J_\alpha\ge1$, so the following changes of Fourier variables preserve the Schwartz class as well. Define the linear bijection $\Theta_F$ by $\Theta_Fe_p=e_{F(p)}$, that is $(\Theta_F\psi)^\wedge(P)=\widehat\psi(F^{-1}P)/J_F(F^{-1}P)$ with $J_F=|\det DF|$, and the product and derivations
\begin{equation}
 \begin{split}
 a\star_Fb&=\Theta_F^{-1}\bigl((\Theta_Fa)(\Theta_Fb)\bigr),\qquad
 e_p\star_Fe_q=e_{F^{-1}(F(p)+F(q))}=e_{p\oplus_Fq},\\
 D_\mu&=\Theta_F^{-1}\partial_\mu\Theta_F,\qquad D_\mu e_p=iF_\mu(p)e_p.
 \end{split}
 \label{s2:star}
\end{equation}
The product is associative and commutative because the pointwise product is, and $D_\mu$ is a derivation of $\star_F$ because $\partial_\mu$ is a derivation of the pointwise product:
\begin{equation}
 D_\mu(a\star_Fb)=\Theta_F^{-1}\partial_\mu\bigl((\Theta_Fa)(\Theta_Fb)\bigr)
 =(D_\mu a)\star_Fb+a\star_F(D_\mu b).
 \label{s2:leibniz}
\end{equation}
On plane waves this is the identity $F_\mu(p\oplus_Fq)=F_\mu(p)+F_\mu(q)$ of Eq.~(5). In the ordinary algebra the same computation gives $\partial_\mu(e_pe_q)=i(p_\mu+q_\mu)e_{p+q}$. The two products have Fourier labels $p+q$ and $p\oplus_Fq$, respectively, whose spatial parts are compared in Theorem~\ref{s2:rigidity}. The derivation $D_\mu$ acts on the latter product with the additive charge $F_\mu(p)+F_\mu(q)$, not with its Fourier label $p\oplus_Fq$. Ordinary convolution in the auxiliary variables followed by $\Theta_F^{-1}$ gives the support formula of the reconstructed algebra,
\begin{equation}
 (a_1\star_F\cdots\star_Fa_n)^\wedge(r)
 =J_F(r)\int\prod_{j=1}^n[\dd p_j]\,(2\pi)^D
 \delta^{(D)}\!\left(F(r)-\sum_{j=1}^nF(p_j)\right)\prod_{j=1}^n\widehat a_j(p_j),
 \label{s2:starsupport}
\end{equation}
so interaction vertices built with $\star_F$ conserve the additive auxiliary momentum $F(p)$. Whenever $F$ is nonlinear the product is nonlocal in the original coordinates. Here locality means that, for each spacetime point $X$, the bilinear distribution kernel is supported at the coincident input positions $(X,X)$. Its Fourier symbol $e^{i(p\oplus_Fq)\cdot X}$ is bounded and hence defines a tempered distribution. A kernel supported at $(X,X)$ is a finite sum of derivatives of delta functions, so its symbol would be $P_X(p,q)e^{i(p+q)\cdot X}$ with $P_X$ polynomial. Unit modulus makes $P_X$ constant, and $q=0$ gives $P_X=1$. By continuity, $(p\oplus_Fq-p-q)\cdot X$ then vanishes: it takes values in $2\pi\mathbb Z$ and is zero at $(0,0)$. Locality for every $X$ therefore forces $p\oplus_Fq=p+q$, or $F(p+q)=F(p)+F(q)$; smoothness makes $F$ linear. Conversely, linear $F$ gives ordinary multiplication. This kernel criterion applies on the stated Fourier test domain without assuming that a general $F$ preserves the Schwartz class.

The isomorphism $\Theta_F$ transports fields and derivations together; nonlocality in the original coordinates concerns the fixed momentum labels and readout conventions, not inequivalence under simultaneous transport of all physical data.

Within this construction, the product and derivations are transported by the same map. If the ordinary pointwise product is kept and only the generators are replaced by $D_\mu$, the Leibniz rule fails, $D_\mu(e_pe_q)=iF_\mu(p+q)e_{p+q}\ne i[F_\mu(p)+F_\mu(q)]e_{p+q}$ for some pairs unless the origin-preserving $F$ is linear; if $\star_F$ is kept with the ordinary derivatives $\partial_\mu$, then $\partial_\mu$ is not a derivation of $\star_F$ for the same reason. Thus the specified map-induced translation sector requires the product and its derivations to be reconstructed compatibly.

For the quartic benchmark, $F=F_\alpha$, the time component is untouched, $D_0=\partial_0$. For the spatial components, $\partial_ie_p=ip_ie_p$ with the covariant component $p_i=-p^i$, and the Laplacian $\Delta=\delta^{ij}\partial_i\partial_j$ gives $\Delta e_p=-|\mathbf p|^2e_p$; hence
\begin{equation}
 D_i=(1-\alpha\Delta)^{1/2}\partial_i,\qquad
 D_ie_p=i\sqrt{1+\alpha|\mathbf p|^2}\,p_ie_p=iF_i(p)e_p,
 \label{s2:quarticderivation}
\end{equation}
where $F_i(p)=\eta_{ij}f_\alpha^j(\mathbf p)$ is the covariant spatial component of $F_\alpha(p)$. The symbol $\sqrt{1+\alpha|\mathbf p|^2}\,p_i$ is not polynomial for $\alpha>0$, so the transported spatial derivations are pseudodifferential rather than finite-order differential operators. The same map fixes the generators, the associative product $\star_\alpha$ with $e_p\star_\alpha e_q=e_{p\oplus_\alpha q}$, and the conservation law used in Sec.~S3.

\begin{revblock}
The composite plane-wave phases mentioned in the main text follow directly from the two product laws. At a common coordinate displacement $X$, their phase-label difference is
\begin{equation}
 \Delta\phi_F(p,q;X)=\bigl(p\oplus_Fq-p-q\bigr)\cdot X.
 \label{s2:phase}
\end{equation}
For $\mathbf p=p_1\mathbf n$, $\mathbf q=p_2\mathbf n$ and $X=(0,L\mathbf n)$, Eq.~\eqref{s2:collinear} and the metric signature give
\begin{equation}
 \Delta\phi_\alpha=\tfrac32\alpha p_1p_2(p_1+p_2)L+O(\alpha^2L).
 \label{s2:collinearphase}
\end{equation}
This is the phase of a composite product, not the interference phase of a one-particle superposition. An observable comparison requires a coupling to that composite product and a common calibration of momenta and displacement; the product identity alone does not specify an interferometer.
\end{revblock}

\subsection{Compton edge position [Eq.~(8)] and numerical bounds}

The process is the head-on collision illustrated in Fig.~1(a): an electron of laboratory energy $E_e$ and spatial momentum $P_{\rm in}\mathbf n$ meets a laser photon of energy $\epsilon_L$ moving along $-\mathbf n$, producing a photon of energy $y_\gamma$ and a recoil electron of momentum $\mathbf p'$. Both descriptions share the quartic shells with the common coefficient $\alpha_{\rm sh} \geq0$,
\begin{equation}
 E^2=m_e^2+|\mathbf p|^2+\alpha_{\rm sh}|\mathbf p|^4\quad(\text{electron}),\qquad
 \omega^2=|\mathbf k|^2+\alpha_{\rm sh}|\mathbf k|^4\quad(\text{photon}),
 \label{s3:shells}
\end{equation}
and, within this kinematic comparison, differ in the prescribed conservation law. The additive (Lorentz-violating) description conserves the Fourier momenta, $p+k=p'+k'$, by Eq.~\eqref{s2:addsupport}. The matched map-induced (DSR) description conserves the auxiliary charges, $F_\alpha(p)+F_\alpha(k)=F_\alpha(p')+F_\alpha(k')$ with $\alpha=\alpha_{\rm sh}$, by Eq.~\eqref{s2:starsupport}; in components,
\begin{equation}
 E_e+\epsilon_L=E'+y_\gamma,\qquad
 \mathbf f_\alpha(\mathbf p)+\mathbf f_\alpha(\mathbf k)
 =\mathbf f_\alpha(\mathbf p')+\mathbf f_\alpha(\mathbf k').
 \label{s3:conservation}
\end{equation}
The endpoints of the photon spectrum in the two descriptions are the quantities $y_{\gamma,\rm add}$ and $y_{\gamma,\Phi}$ of Fig.~1(b). The incoming energies $E_e$ and $\epsilon_L$ are the laboratory inputs held fixed in the comparison. Their common preparation and calibration, and the common detector response introduced below, are assumptions of the comparison; they are not inferred from the shell map. The calculation determines kinematically allowed endpoints; their experimental population also requires a nonvanishing scattering rate near the endpoint. The spectrum curves in Fig.~1(b) are schematic: no spectral shape or event rate is derived here.

For $\alpha_{\rm sh}=0$ let $P_0=\sqrt{E_e^2-m_e^2}$. A photon emitted at angle $\theta$ to $\mathbf n$ leaves the recoil electron with energy $E_e+\epsilon_L-y$ and momentum $\mathbf p_{\rm in}+\mathbf k-\mathbf k'$; its mass shell gives
\begin{equation}
 y(\theta)=\frac{\epsilon_L(E_e+P_0)}{E_e+\epsilon_L-(P_0-\epsilon_L)\cos\theta},
 \label{s3:angular}
\end{equation}
which for $P_0>\epsilon_L$ (the inverse-Compton regime) is maximal at $\theta=0$. The Compton edge is
\begin{equation}
 y_0=\frac{\epsilon_L(E_e+P_0)}{D},\qquad
 D\equiv E_e-P_0+2\epsilon_L=\frac{m_e^2}{E_e+P_0}+2\epsilon_L,
 \label{s3:y0}
\end{equation}
with recoil-electron variables $P'_0=P_0-\epsilon_L-y_0$, $E'_0=E_e+\epsilon_L-y_0$, so that $E'_0-P'_0=D$ and $v'_0\equiv P'_0/E'_0$. The factor $E_e+P_0$ in the numerator and the small recoil denominator $D$ are the origin of the large lever arm found below.

\begin{proposition}\label{s3:matched}
In the matched map-induced description the kinematic Compton edge is exactly $y_{\gamma,\rm DSR}^{\rm match}=y_0$ for every $\alpha_{\rm sh}\geq0$.
\end{proposition}
\begin{proof}
Apply $F_\alpha$ to all momenta. By Eq.~\eqref{s1:shellidentity}, the auxiliary momenta of the electron and of the photon lie on the ordinary shells of masses $m_e$ and $0$ with the same energies, since $F_\alpha$ preserves the energy; by Eq.~\eqref{s3:conservation} they obey ordinary additive conservation; and since $\mathbf f_\alpha$ is radial, the directions of all momenta are preserved. In the auxiliary variables the process is therefore ordinary Compton scattering of an electron of energy $E_e$ with a photon of energy $\epsilon_L$ in head-on geometry, whose maximal photon energy is $y_0$ by Eq.~\eqref{s3:y0}. The photon energy is unchanged by $F_\alpha$, so the maximal value of $y_\gamma$ is $y_0$.
\end{proof}

In the additive description the edge is obtained to first order as follows. All expansions below are taken as $\alpha_{\rm sh}\to0$ at fixed finite incoming energies, on the branch continuously connected to the undeformed endpoint. At that endpoint the recoil-shell constraint has derivative $-2D\ne0$ with respect to $y_\gamma$, ensuring a smooth local branch. At fixed energies, the Fourier momenta on the quartic shells \eqref{s3:shells} are smaller than the auxiliary ones. Solving $E^2=m^2+P^2+\alpha_{\rm sh}P^4$ for $P$,
\begin{equation}
\begin{aligned}
 P_{\rm in}&=P_0-\frac{\alpha_{\rm sh}}2P_0^3+O(\alpha_{\rm sh}^2),&
 \kappa&=\epsilon_L-\frac{\alpha_{\rm sh}}2\epsilon_L^3+O(\alpha_{\rm sh}^2),\\
 K&=y_\gamma-\frac{\alpha_{\rm sh}}2y_\gamma^3+O(\alpha_{\rm sh}^2),
\end{aligned}\label{s3:fixedenergy}
\end{equation}
for the incoming electron, the laser photon and the outgoing photon respectively. At the endpoint all momenta are collinear; additive conservation along $\mathbf n$ reads $P_{\rm in}-\kappa=P'+K$, energy conservation gives $E'=E_e+\epsilon_L-y_\gamma$, and the recoil electron is on its shell, $E'^2=m_e^2+P'^2+\alpha_{\rm sh}P'^4$. Writing $y_\gamma=y_0+\alpha_{\rm sh}\delta y$ and $P'=P'_0+\alpha_{\rm sh}\delta P'$, momentum conservation and Eq.~\eqref{s3:fixedenergy} give
\begin{equation}
 \delta P'=-\delta y-\tfrac12(P_0^3-\epsilon_L^3-y_0^3),
 \label{s3:deltap}
\end{equation}
and the recoil shell, expanded to first order, gives $-E'_0\delta y=P'_0\delta P'+\tfrac12P_0'^4$. Eliminating $\delta P'$,
\begin{equation}
\begin{aligned}
 (E'_0-P'_0)\delta y&=P'_0S,\\
 S&\equiv\tfrac12(P_0^3-\epsilon_L^3-y_0^3-P_0'^3)
 =\tfrac32\bigl[P_0\epsilon_L(P_0-\epsilon_L)+P'_0y_0(P'_0+y_0)\bigr],
\end{aligned}\label{s3:source}
\end{equation}
where the second form of $S$ follows from $P_0-\epsilon_L=P'_0+y_0$ by factoring the cubic differences. Since $E'_0-P'_0=D$, the additive endpoint is
\begin{equation}
 y_{\gamma,\rm LIV}=y_0-\alpha_{\rm sh}C_\gamma+O(\alpha_{\rm sh}^2),\qquad
 C_\gamma\equiv-\frac{P'_0}{D}S=-\frac{v'_0}{1-v'_0}S.
 \label{s3:Cgamma}
\end{equation}
For $P_0>\epsilon_L$ and forward recoil $P'_0>0$, $S>0$ and $C_\gamma<0$: the additive endpoint lies above the matched one, $y_{\gamma,\rm add}>y_{\gamma,\Phi}$ at positive sufficiently small $\alpha_{\rm sh}$. The collinear configuration suffices at first order: for the undeformed kinematics the emission angle $\theta=0$ is a nondegenerate maximum of Eq.~\eqref{s3:angular}, so a first-order displacement of the maximizing direction changes the maximal value only at second order.

The result \eqref{s3:Cgamma} can be obtained directly from the difference between the two conservation laws, which exhibits the compensation described in the main text. For collinear momenta, $\mathbf f_\alpha(P\mathbf n)=(P+\tfrac12\alpha_{\rm sh}P^3+\cdots)\mathbf n$, so the matched spatial law \eqref{s3:conservation} expanded to first order reads
\begin{equation}
 P_{\rm in}-\kappa=P'+K-\alpha_{\rm sh}S+O(\alpha_{\rm sh}^2),
 \label{s3:compensation}
\end{equation}
where $S$ is exactly the difference of the collinear defects \eqref{s2:collinear} of the initial and final pairs, $S=\xi(P_0,-\epsilon_L)-\xi(P'_0,y_0)$ with $\xi(a,b)=-\tfrac32ab(a+b)$. Subtracting the additive law $P_{\rm in}-\kappa=P'+K$, which has the same incoming momenta, gives $(\delta P'_{\rm DSR}-\delta P'_{\rm LIV})+(\delta y_{\rm DSR}-\delta y_{\rm LIV})=S$. The recoil shell is the same in both descriptions, so $\delta P'_{\rm DSR}-\delta P'_{\rm LIV}=-(\delta y_{\rm DSR}-\delta y_{\rm LIV})/v'_0$, and therefore
\begin{equation}
 \delta y_{\rm DSR}-\delta y_{\rm LIV}=-\frac{v'_0}{1-v'_0}S=C_\gamma.
 \label{s3:separation}
\end{equation}
With $\delta y_{\rm DSR}=0$ from Proposition~\ref{s3:matched}, Eq.~\eqref{s3:separation} reproduces Eq.~\eqref{s3:Cgamma}. The common shell correction \eqref{s3:fixedenergy} yields the additive shift $-\alpha_{\rm sh}C_\gamma$; the matched-law term $-\alpha_{\rm sh}S$ in Eq.~\eqref{s3:compensation} compensates it. Under the stated common-input assumptions, the endpoint separation therefore isolates the composition-law difference.

Let the measured edge position be $y=M(y_\gamma)$, where $M$ is the common, independently calibrated response of the apparatus (a smooth increasing function for an energy-calibrated photon channel). Define
\begin{equation}
 C_{\rm op}=M'(y_0)C_\gamma,\quad y_{\rm SR}=M(y_0),\quad
 y_{\rm LIV}=M(y_{\gamma,\rm LIV}),\quad
 y_{\rm DSR}^{\rm match}=M(y_{\gamma,\rm DSR}^{\rm match}).
 \label{s3:response}
\end{equation}
Then Eq.~\eqref{s3:Cgamma} and Proposition~\ref{s3:matched} give
\begin{equation}
\begin{aligned}
 y_{\rm LIV}&=y_{\rm SR}-\alpha_{\rm sh}C_{\rm op}+O(\alpha_{\rm sh}^2),\qquad
 y_{\rm DSR}^{\rm match}=y_{\rm SR},\\
 |y_{\rm LIV}-y_{\rm DSR}^{\rm match}|&=|y_{\rm LIV}-y_{\rm SR}|
 =|\alpha_{\rm sh}C_{\rm op}|+O(\alpha_{\rm sh}^2),
\end{aligned}\label{s3:equalorder}
\end{equation}
which is Eq.~(8) for the photon-energy endpoint of Fig.~1(b). For an energy-calibrated photon channel, $y=y_\gamma$ and $C_{\rm op}=C_\gamma$. The coefficient $C_{\rm op}$ depends on the beam energy and the laser frequency and the collision geometry through $C_\gamma$, and on the detector response through $M'$. If the edge is read instead as the position $X=H(P')$ of the recoil electron in a spectrometer, with the same smooth response $H$ and $P'_0>0$, the common recoil shell gives $\delta P'_{\rm DSR}-\delta P'_{\rm LIV}=-C_\gamma/v'_0$, so that the additive and matched endpoints are separated by $|X_{\rm LIV}-X_{\rm DSR}^{\rm match}|=\alpha_{\rm sh}|H'(P'_0)C_\gamma/v'_0|+O(\alpha_{\rm sh}^2)$. Both recoil momenta contain the common shell contribution $-\tfrac12\alpha_{\rm sh}P_0'^3$. Thus the raw spectrometer positions do not satisfy Eq.~(8) relative to the SR position: the common shell contribution must first be removed through an independent shell calibration. The remaining separation is the composition-law contribution.

The coefficient also incorporates a nonzero collision geometry. For a crossing-angle deviation $\chi$ from head-on incidence, let $\mathbf n$ and $\bm\ell$ be the unit directions of the electron and of the laser photon, $\mathbf n\cdot\bm\ell=-\cos\chi$, and set
\begin{equation}
 \mathbf S_0=P_0\mathbf n+\epsilon_L\bm\ell,\quad q_0=|\mathbf S_0|,\quad
 \mathbf u=\mathbf S_0/q_0,\quad T_0=E_e+\epsilon_L,\quad s_0=T_0^2-q_0^2.
 \label{s3:geometry}
\end{equation}
The undeformed photon energy at angle $\theta$ to $\mathbf u$ is $y(\theta)=(s_0-m_e^2)/[2(T_0-q_0\cos\theta)]$, maximal at $\theta=0$, so that
\begin{equation}
 y_0(\chi)=\frac{(s_0-m_e^2)(T_0+q_0)}{2s_0},\qquad
 P'_0=q_0-y_0(\chi),\qquad E'_0-P'_0=T_0-q_0,
 \label{s3:crossedge}
\end{equation}
in the forward-recoil regime $q_0>y_0(\chi)$. Repeating the computation leading to Eq.~\eqref{s3:Cgamma} with vector momenta, the first-order angular displacement of the endpoint direction again drops out, and the composition source is the projection on $\mathbf u$ of the difference of the defects of the initial and final pairs,
\begin{equation}
\begin{aligned}
 S(\chi)&=\mathbf u\cdot\bigl[\bm\Xi^{(1)}(P_0\mathbf n,\epsilon_L\bm\ell)
 -\bm\Xi^{(1)}(P'_0\mathbf u,y_0\mathbf u)\bigr]\\
 &=\tfrac12\bigl[P_0^3(\mathbf n\cdot\mathbf u)+\epsilon_L^3(\bm\ell\cdot\mathbf u)-P_0'^3-y_0^3\bigr],\qquad
 C_\gamma(\chi)=-\frac{P'_0}{T_0-q_0}S(\chi).
\end{aligned}\label{s3:crosssource}
\end{equation}
Proposition~\ref{s3:matched} holds verbatim, since $F_\alpha$ preserves all directions. The head-on case $\bm\ell=-\mathbf n$, $q_0=P_0-\epsilon_L$, $\mathbf u=\mathbf n$ reproduces Eqs.~\eqref{s3:y0}-\eqref{s3:Cgamma}.

\begin{figure}[tbp]
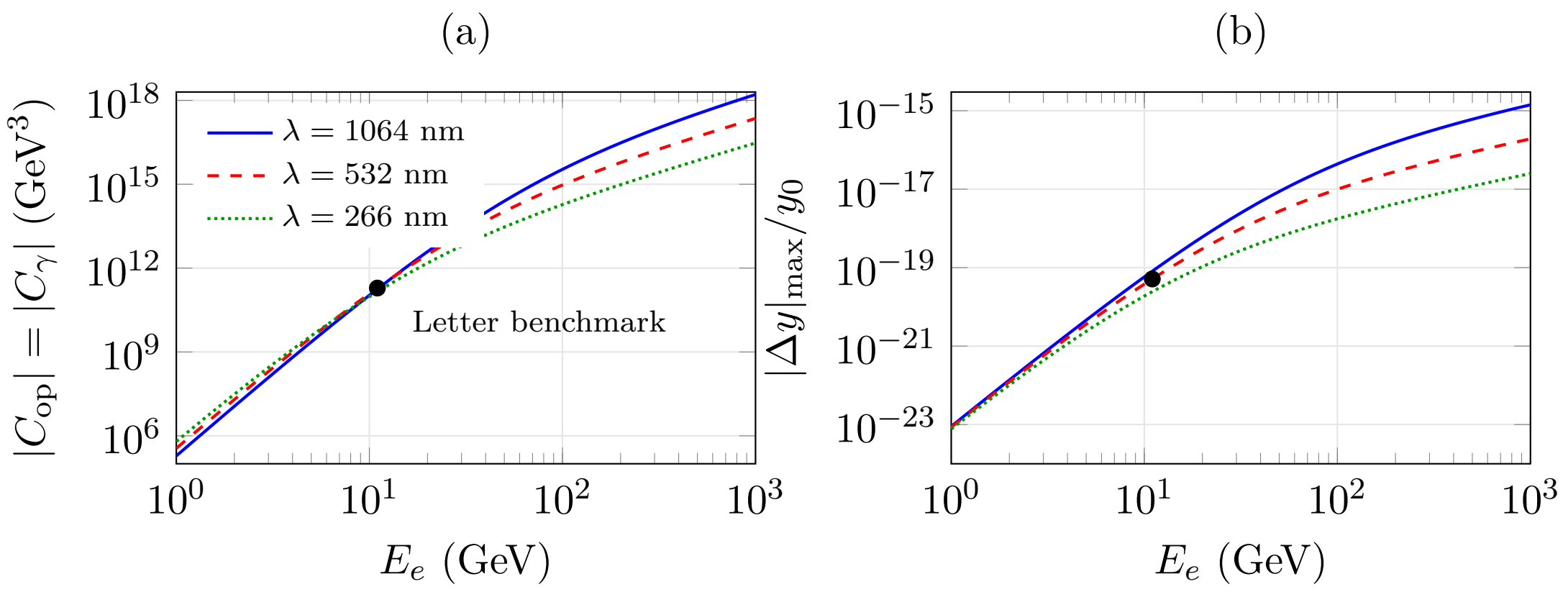

\centering
\comptonplot
\caption{Head-on Compton-edge scaling from Eqs.~\eqref{s3:y0}-\eqref{s3:Cgamma}. (a) $|C_{\rm op}|=|C_\gamma|$ for an energy-calibrated photon channel and three laser wavelengths. (b) Leading relative separation $\alpha_{\rm sh}|C_\gamma|/y_0$ at the rounded limiting coefficient for the additive realization, $\alpha_{\rm sh}=8.25\times10^{-31}\,\mathrm{GeV}^{-2}$. The dot marks $E_e=11\,\mathrm{GeV}$, $\lambda=532\,\mathrm{nm}$. The high-energy approximation \eqref{s3:scaling} explains the crossing in (a) and the decrease with laser frequency in (b).}
\label{s3:figure}
\end{figure}

The head-on, high-energy form gives the kinematic coefficient $C_\gamma$. With the Compton parameter $x=4E_e\epsilon_L/m_e^2$ and $E_e\gg m_e$, for laser energies $\epsilon_L\ll m_e$, Eqs.~\eqref{s3:y0}-\eqref{s3:Cgamma} reduce to $y_0\simeq E_ex/(1+x)$, $P'_0\simeq E_e/(1+x)$, $D\simeq m_e^2(1+x)/(2E_e)$ and $S\simeq\tfrac32 E_e^3x/(1+x)^2$, hence
\begin{equation}
 C_\gamma\simeq-\frac{3E_e^5x}{m_e^2(1+x)^4},\qquad
 \frac{|\alpha_{\rm sh}C_\gamma|}{y_0}\simeq\alpha_{\rm sh}\frac{3E_e^4}{m_e^2(1+x)^3}.
 \label{s3:scaling}
\end{equation}
Within this approximation, at fixed laser frequency $x\propto E_e$, giving $|C_\gamma|\propto E_e^6$ for $x\ll1$ and $|C_\gamma|\propto E_e^2$ for $x\gg1$. At fixed beam energy, $x/(1+x)^4$ is maximal at $x=1/3$ and explains the curve crossings in Fig.~\ref{s3:figure}(a), which plots the exact coefficient for three wavelengths. These scalings apply to an energy-calibrated photon channel; generally $C_{\rm op}=M'(y_0)C_\gamma$. Measurements across calibrated beam energies and laser frequencies could test this scaling against a constant offset, but not arbitrary energy-dependent systematic effects.

To obtain the astrophysical bound quoted in the main text, consider additive photon decay $\gamma\to e^+e^-$. Write $\omega(k)^2=k^2+\alpha_\gamma k^4$ for the photon and $E_e(r)^2=r^2+m_e^2+\alpha_e r^4$ for the electron, with $k=|\mathbf k|$ and $r=|\mathbf p|$.

\Needspace{7\baselineskip}
\begin{lemma}\label{s3:convexity}
For $\alpha_e\geq0$ the function $\mathbf p\mapsto E_e(|\mathbf p|)$ is strictly convex on $\mathbb R^d$.
\end{lemma}
\begin{proof}
The Hessian has the tangential eigenvalue $E'_e(r)/r=(1+2\alpha_er^2)/E_e(r)>0$ and the radial eigenvalue $E_e''(r)=[m_e^2(1+6\alpha_er^2)+3\alpha_er^4+2\alpha_e^2r^6]/E_e(r)^3>0$. At $r=0$ both expressions have the positive limit $1/m_e$.
\end{proof}

For a decay $\mathbf k\to\mathbf p_++\mathbf p_-$ with $\mathbf p_\pm=\mathbf k/2\pm\mathbf v$, convexity gives $E_e(|\mathbf p_+|)+E_e(|\mathbf p_-|)\geq2E_e(k/2)$ with equality at $\mathbf v=0$, so the decay is kinematically allowed if and only if $\omega(k)\geq2E_e(k/2)$. Squaring the threshold equality,
\begin{equation}
 \left(\alpha_\gamma-\frac{\alpha_e}{4}\right)k^4=4m_e^2,
 \label{s3:threshold}
\end{equation}
so a finite threshold exists only for $\alpha_\gamma>\alpha_e/4$. On the common shell, $\alpha_\gamma=\alpha_e=\alpha_{\rm sh}>0$, the threshold is $k_{\rm th}=[16m_e^2/(3\alpha_{\rm sh})]^{1/4}$. Assuming a decay coupling for which the decay is prompt above threshold on the propagation baseline, compatibility with a lower limit $E_{\rm cut}$ on the cutoff energy requires $E_{\rm th}=\omega(k_{\rm th})>E_{\rm cut}$. Here $E_{\rm th}^2=k_{\rm th}^2+16m_e^2/3$, so replacing the threshold momentum by the photon energy incurs a relative correction of order $m_e^2/E_{\rm cut}^2$. To this accuracy,
\begin{equation}
 \alpha_{\rm sh}<\frac{16m_e^2}{3E_{\rm cut}^4},\qquad
 \Lambda_\alpha\equiv\alpha_{\rm sh}^{-1/2}>
 \left(\frac{3E_{\rm cut}^4}{16m_e^2}\right)^{1/2}.
 \label{s3:bound}
\end{equation}
With the nominal $95\%$ confidence-level lower limit $E_{\rm cut}=1.14\,\mathrm{PeV}$ on the photon cutoff energy reported by the LHAASO analysis of J2032+4102 \cite{Cao2022LHAASO} and $m_e=0.51099895\times10^{-3}\,\mathrm{GeV}$,
\begin{equation}
 \alpha_{\rm sh}<8.24555\times10^{-31}\,\mathrm{GeV}^{-2},\qquad
 \Lambda_\alpha>1.10126\times10^{15}\,\mathrm{GeV},
 \label{s3:boundnumbers}
\end{equation}
which we quote as $\alpha<8.25\times10^{-31}\,\mathrm{GeV}^{-2}$ and $\Lambda_\alpha>1.10\times10^{15}\,\mathrm{GeV}$. Table~I of Ref.~\cite{Cao2022LHAASO} also reports a systematic variation of $\pm0.06\,\mathrm{PeV}$ in this cutoff limit. The numerical bounds here and in the main text use its nominal value; the additional digits document the calculation, not experimental precision. The corresponding upper limits on $\alpha_{\rm sh}$ and the leading endpoint separation scale as $E_{\rm cut}^{-4}$, while the lower limit on $\Lambda_\alpha$ scales as $E_{\rm cut}^{2}$. This systematic variation is separate from the quoted confidence level. Equation~\eqref{s3:bound} is the threshold appropriate to the common photon-electron shell; the photon-only threshold ($\alpha_e=0$) would give instead $\alpha_\gamma<4m_e^2/E_{\rm cut}^4$ to the same accuracy. Threshold reactions of this type are reviewed in Ref.~\cite{PhysRevD.67.124011}.

\begin{table}[tbp]
\centering
\caption{Head-on Compton-edge quantities for a $532\,\mathrm{nm}$ laser ($\epsilon_L=2.33053\,\mathrm{eV}$), from Eqs.~\eqref{s3:y0}-\eqref{s3:Cgamma}. The last two columns give the leading separation at the rounded additive-model limit, $\alpha_{\rm sh}=8.25\times10^{-31}\,\mathrm{GeV}^{-2}$.}
\label{s3:table}
\small
\begin{tabular}{rrrrrr}
\toprule
$E_e\ (\mathrm{GeV})$ & $x$ & $y_0\ (\mathrm{GeV})$ & $C_\gamma\ (\mathrm{GeV}^3)$ & $|\Delta y|_{\max}\ (\mathrm{eV})$ & $|\Delta y|_{\max}/y_0$\\
\midrule
6.03 & 0.215 & 1.068 & $-9.04\times10^9$ & $7.46\times10^{-12}$ & $6.98\times10^{-21}$\\
11 & 0.393 & 3.102 & $-1.93\times10^{11}$ & $1.59\times10^{-10}$ & $5.14\times10^{-20}$\\
20 & 0.714 & 8.331 & $-3.04\times10^{12}$ & $2.51\times10^{-9}$ & $3.01\times10^{-19}$\\
50 & 1.785 & 32.05 & $-1.07\times10^{14}$ & $8.79\times10^{-8}$ & $2.74\times10^{-18}$\\
100 & 3.570 & 78.12 & $-9.40\times10^{14}$ & $7.76\times10^{-7}$ & $9.93\times10^{-18}$\\
250 & 8.925 & 224.8 & $-1.03\times10^{16}$ & $8.51\times10^{-6}$ & $3.79\times10^{-17}$\\
\bottomrule
\end{tabular}
\end{table}

In the matched map-induced description the same shell gives no threshold at all: the auxiliary momenta obey ordinary conservation, and a future-directed null vector cannot equal the sum of two future-directed timelike vectors of mass $m_e$, since $(P_++P_-)^2=2m_e^2+2P_+\cdot P_-\geq4m_e^2>0$. The astrophysical bound therefore constrains the additive realization of the common shell. It does not independently bound the matched realization. Restricting the common coefficient to the range allowed for the additive model, Eq.~(8) gives the maximal leading endpoint separation between the two realizations in that range.

For the benchmark discussed in the main text, $E_e=11\,\mathrm{GeV}$ and $\lambda=532\,\mathrm{nm}$, $\epsilon_L=hc/\lambda=2.33053\,\mathrm{eV}$, and Eqs.~\eqref{s3:y0}-\eqref{s3:Cgamma} give
\begin{equation}
\begin{aligned}
 E_e-P_0&=11.869\,\mathrm{eV},& D&=16.530\,\mathrm{eV},&y_0&=3.10171\,\mathrm{GeV},\\
 P'_0&=7.89829\,\mathrm{GeV},&P'_0/D&=4.7781\times10^8,&S&=404.22\,\mathrm{GeV}^3,
\end{aligned}\label{s3:benchmark}
\end{equation}
hence $C_\gamma=-1.93141\times10^{11}\,\mathrm{GeV}^3$ (the high-energy form \eqref{s3:scaling} reproduces this value to all digits shown, with $x=0.3927$). At the rounded limiting coefficient $\alpha_{\rm sh}=8.25\times10^{-31}\,\mathrm{GeV}^{-2}$ used in the figure and table, the leading endpoint separation and its relative value are approximately $1.5934\times10^{-10}\,\mathrm{eV}$ and $5.1372\times10^{-20}$. Using the unrounded astrophysical limit from Eq.~\eqref{s3:bound} gives the strict rounded bounds quoted in the main text,
\begin{revblock}
\begin{equation}
 \begin{aligned}
 |\Delta y_{\rm LIV-SR}|_{\rm leading}
 =|\Delta y_{\rm LIV-DSR}^{\rm match}|_{\rm leading}
 &=\alpha_{\rm sh}|C_\gamma|<1.593\times10^{-10}\,\mathrm{eV},\\
 \frac{|\Delta y|_{\rm leading}}{y_0}&<5.14\times10^{-20}.
 \end{aligned}
 \label{s3:finalnumbers}
\end{equation}
\end{revblock}
Table~\ref{s3:table} lists the same quantities for other beam energies at the same laser wavelength. The separation is far below the precision of existing edge determinations \cite{Bocquet2010Compton,Gharibyan2012}, and the table and Fig.~\ref{s3:figure} quantify the increase in the kinematic response at higher beam energy. Improved bounds would additionally require sufficient measurement precision and control of systematic effects.

\bibliography{refs}

\end{document}